\documentclass[journal]{IEEEtran}

\usepackage{cite}
\usepackage{tcolorbox}
\usepackage{fontawesome}
\usepackage{gensymb}
\usepackage{algorithmic}
\usepackage{algorithm}
\usepackage{bm}
\usepackage{booktabs}       % professional-quality tables
\usepackage{subcaption}
\usepackage{nicefrac}
\usepackage{graphicx}
\usepackage{textcomp}
\usepackage{xcolor}
\usepackage{balance}
\usepackage{tikz}
\usetikzlibrary{arrows.meta,positioning}
\def\BibTeX{{\rm B\kern-.05em{\sc i\kern-.025em b}\kern-.08em
    T\kern-.1667em\lower.7ex\hbox{E}\kern-.125emX}}
\usepackage{url}            % simple URL typesetting
\usepackage{booktabs}       % professional-quality tables
\usepackage{nicefrac}       % compact symbols for 1/2, etc.
\usepackage{microtype}      % microtypography
\usepackage{lipsum}
\usepackage{amsmath,amsthm,amsfonts,amssymb,amscd}
\usepackage{verbatim}
\usepackage{dsfont}
\usepackage{mdframed}
\usepackage{empheq}

\newtheorem{defi}{Definition}
\newtheorem{theo}{Theorem}

\title{Fast and Secure Routing Algorithms for Quantum Key Distribution Networks}
\author{
\IEEEauthorblockN{Shahbaz Akhtar\IEEEauthorrefmark{1}, 
Krishnakumar G\IEEEauthorrefmark{2}, Vishnu B\IEEEauthorrefmark{3}, and Abhishek Sinha \IEEEauthorrefmark{4}\\}
\IEEEauthorblockA{\IEEEauthorrefmark{1}\IEEEauthorrefmark{2}\IEEEauthorrefmark{3}Department of Electrical Engineering,  Indian Institute of Technology Madras, Chennai 600036, India\\
\IEEEauthorrefmark{4} School of Technology and Computer Science, Tata Institute of Fundamental Research, Mumbai 400 005, India }
Email: 
\IEEEauthorrefmark{1}akhtar.dr.shahbaz@gmail.com,
\IEEEauthorrefmark{2} krishnakumar97@smail.iitm.ac.in,
\IEEEauthorrefmark{3} vishnubeji@gmail.com,
\IEEEauthorrefmark{4} abhishek.sinha@tifr.res.in
\thanks{This paper was presented in part at  \cite{QKD-conf}.}%.
}

\begin{document}
\maketitle
\begin{abstract}
We consider the problem of secure packet routing at the maximum achievable rate in Quantum Key Distribution (QKD) networks. Assume that a QKD protocol generates symmetric private key pairs for secure communication over each link in a network. The quantum key generation process is modeled using a stochastic counting process. Packets are first encrypted with the quantum keys available for each hop and then transmitted on a point-to-point basis over the links. A fundamental problem in this setting is the design of a secure and capacity-achieving routing policy that takes into account the time-varying availability of the encryption keys and finite link capacities. In this paper, we propose a new secure throughput-optimal policy called \emph{Tandem Queue Decomposition} (TQD). The TQD policy is derived by combining the QKD process with the Universal Max Weight routing policy, proposed earlier by Sinha and Modiano. We show that the TQD policy solves the problem of secure and efficient packet routing for a broad class of traffic, including unicast, broadcast, multicast, and anycast. The proposed decomposition reduces the problem to the generalized network flow problem without the key availability constraints over a transformed network. The proof of the  throughput-optimality of the TQD policy uses the Lyapunov stability theory for analyzing the interdependent packet queueing process and the key-storage dynamics. Finally, we demonstrate the competitiveness of the TQD policy over the existing algorithms by numerically comparing them on a simulator that we build on top of the state-of-the-art OMNeT++ network simulator platform.
\end{abstract}
\begin{IEEEkeywords}
Quantum Key Distribution, Throughput-optimal routing, Network Algorithms.	
\end{IEEEkeywords}
\section{Introduction} \label{secA}
\begin{figure}[h]
\centering\includegraphics[width=0.48\textwidth]{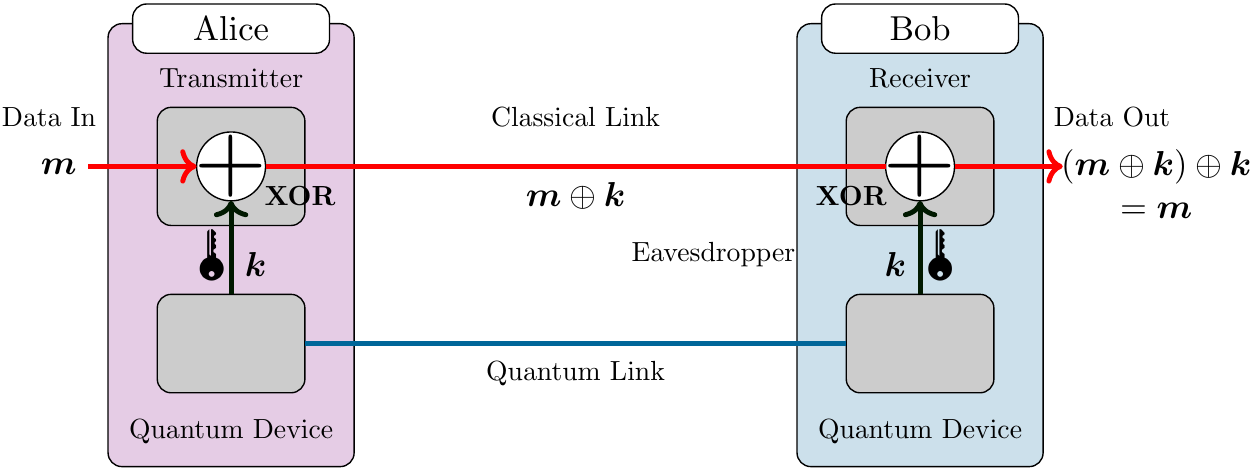}
\put(-119,44){\includegraphics[width=0.04\textwidth]{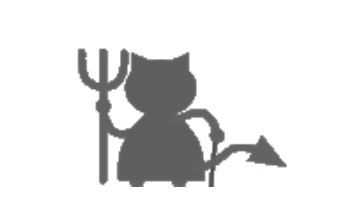}}
%\includegraphics[scale=0.1]{demon-cropped-new.pdf}
%\put(-238, 60){\footnotesize{$\bm{m}$}}
%\put(-189,52){\footnotesize{$\bm{k}$}}
%\put(-134, 65){\footnotesize{$\bm{m} \oplus \bm{k}$}}
%\put(-38,52){\footnotesize{$\bm{k}$}}
%\put(-16, 60){\footnotesize{$(\bm{m}\oplus \bm{k})\oplus \bm{k}$}}
%\put(-16, 52){\footnotesize{$=\bm{m}$}}
%\put(-111, 52){\scriptsize{Eavesdropper}}
\caption{\footnotesize{Depicting a QKD link with the One-Time Pad (OTP) encryption protocol. A sufficiently long symmetric encryption key $\bm{k}$ is first established between Alice and Bob via the Quantum Link using Quantum Entanglement mechanisms. Next, the message $\bm{m}$ from Alice to Bob is encrypted at the source by taking XOR of the message with the shared key (bit-by-bit). The encrypted message $\bm{m} \oplus \bm{k}$ is then transmitted over the classical link. Upon receiving the encrypted message, Bob securely decrypts it by XORing it against the same shared secret key as $(\bm{m}\oplus \bm{k})\oplus \bm{k} =\bm{m}$. The eavesdropper may try to peek at the message transmitted over the classical link.}}
 \label{linkin1}
\end{figure}

\IEEEPARstart{Q}{uantum} key distribution (QKD) enables two geographically separate communicating parties to exchange symmetric private keys, whose information-theoretical security is guaranteed by the fundamental principles of quantum mechanics \cite{lo1999unconditional, shor2000simple, van2014quantum}. The generated private keys are used for encrypting messages that are communicated over the classical channels (\emph{e.g.,} free space or optical fibers). Many QKD protocols are known and are already in use, including BB84 \cite{shor2000simple}, E91 \cite{ling2008experimental}, and B92 \cite{sasaki2015key}. QKD protocols use quantum effects, such as the no-cloning property and quantum entanglement, to detect possible eavesdropping by an adversarial third party. Once the peer nodes have mutually established secret keys, the messages exchanged between them can be securely encrypted using standard symmetric ciphers, such as One-time Pad (OTP) or Advanced Encryption Standard (AES) \cite{elliott2002building, websites}. We emphasize that QKD is used only for establishing the secret keys; the encrypted messages are transmitted exclusively over the classical links. QKD schemes should be contrasted with the ongoing research on Post-Quantum Cryptography (PQC) that, although believed to be secure against attack with quantum computers, lacks formal guarantees for their security properties \cite{bernstein2017post}. Much progress has recently been made in the practical implementations of various QKD schemes \cite{peev2009secoqc, tang2018quantum, evans}. 
%In this paper, we study a QKD system similar to the one recently implemented by a team from Oak Ridge and Los Alamos National Labs \cite{evans}. 
See Figure \ref{linkin1} for a schematic of a one-hop QKD system.

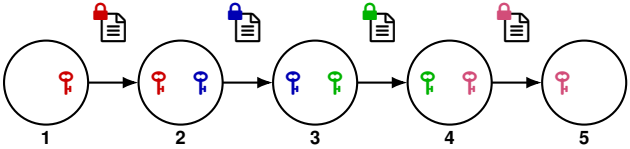
\begin{figure}
    \centering
    \resizebox{0.47\textwidth}{!}{\colorlet{dRed}{red!80!black}
\colorlet{dBlue}{blue!70!black}
\colorlet{dGreen}{green!70!black}
\colorlet{dPink}{purple!70!white}

\def\keyht{3cm}
\def\antr{0.75cm}
\def\ndr{4cm}
\def\yofst{5.25cm}
\def\xfctr{3.2}
\def\tx{2cm}
\def\yantofstfctr{0.45}
\tikzset{
  redKey/.pic={
    code={
      \filldraw [line width=2mm, dRed] (0,0) -- (0,\keyht);
      \fill [even odd rule, inner color=white, outer color=dRed] (0,\keyht) circle (\antr);
    }
  },
  blueKey/.pic={
    code={
      \filldraw [line width=2mm, dBlue] (0,0) -- (0,\keyht);
      \fill [even odd rule, inner color=white, outer color=dBlue] (0,\keyht) circle (\antr);
    }
  },
  greenKey/.pic={
    code={
      \filldraw [line width=2mm, dGreen] (0,0) -- (0,\keyht);
      \fill [even odd rule, inner color=white, outer color=dGreen] (0,\keyht) circle (\antr);
    }
  },
  pinkKey/.pic={
    code={
      \filldraw [line width=2mm, dPink] (0,0) -- (0,\keyht);
      \fill [even odd rule, inner color=white, outer color=dPink] (0,\keyht) circle (\antr);
    }
  }
}

\begin{tikzpicture}
\draw [line width=2mm] (0,0) circle (\ndr);
\node[scale=6] at (0.5*\ndr,0) {\color{dRed}\rotatebox{-45}{\faKey}};

\draw [line width=2mm] (\xfctr*\ndr,0) circle (\ndr);
\node[scale=6] at (\xfctr*\ndr-0.5*\ndr,0) {\color{dRed}\sffamily\rotatebox{-45}{\faKey}};
\node[scale=6] at (\xfctr*\ndr+0.5*\ndr,0) {\color{dBlue}\rotatebox{-45}{\faKey}};

\draw [line width=2mm] (2*\xfctr*\ndr,0) circle (\ndr);
\node[scale=6] at (2*\xfctr*\ndr-0.5*\ndr,0) {\color{dBlue}\rotatebox{-45}{\faKey}};
\node[scale=6] at (2*\xfctr*\ndr+0.5*\ndr,0) {\color{dGreen}\rotatebox{-45}{\faKey}};

\draw [line width=2mm] (3*\xfctr*\ndr,0) circle (\ndr);
\node[scale=6] at (3*\xfctr*\ndr-0.5*\ndr,0) {\color{dGreen}\rotatebox{-45}{\faKey}};
\node[scale=6] at (3*\xfctr*\ndr+0.5*\ndr,0) {\color{dPink} \rotatebox{-45}{\faKey}};

\draw [line width=2mm] (4*\xfctr*\ndr,0) circle (\ndr);
\node[scale=6] at (4*\xfctr*\ndr-0.5*\ndr,0) {\color{dPink}\rotatebox{-45}{\faKey}};

\foreach \i in {0,...,4}{
\pgfmathtruncatemacro{\n}{\i+1}
\node[anchor=center,scale=5] at (\i*\xfctr*\ndr,-\yofst) {\sffamily\bfseries\n};
}

\foreach \i/\j in {1/dRed,2/dBlue,3/dGreen,4/dPink}{
\node[anchor=center,scale=8] at (\i*\xfctr*\ndr-1.6*\ndr,\yofst) {\sffamily\faFileTextO};
\node[anchor=center,scale=6] at (\i*\xfctr*\ndr-1.9*\ndr,\yofst+1.5cm) {\sffamily\color{\j}\faLock};
}

\foreach \i in {1,...,4}{
\pgfmathtruncatemacro{\p}{\i-1}
\draw [-{Latex[scale=1.5]}, line width=2mm] (\p*\xfctr*\ndr+\ndr,0) -- (\i*\xfctr*\ndr-\ndr,0);
}
\end{tikzpicture}}
    \caption{\small{Illustrating the Trusted-node QKD Architecture}}
    \label{tnode}
\end{figure}

Despite having excellent security properties, the basic QKD scheme is severely distance-limited and requires the use of quantum repeaters to enable long-distance communication via the entanglement swapping mechanism \cite{dur1999quantum}. Unfortunately, due to the difficulty in fabricating short-term quantum memories, quantum repeaters have so far been proven infeasible to build in a scalable and cost-effective fashion \cite{yan2021survey}. These limitations can be largely mitigated by building QKD networks with stand-alone QKD links.
In this paper, we focus on the widely used ``Trusted Node/Relay" setup, where each communication link is assumed to be equipped with a dedicated QKD channel with secure endpoints \cite{evans, alleaume2014using, peev2009secoqc, sasaki2015key, stucki2011long, elliott2002building}. See Figure \ref{tnode} for a schematic. In this architecture, each packet is sequentially encrypted and decrypted along its path by the trusted nodes on each of the intermediate hops. The transmitted messages on each link are encrypted to prevent the eavesdropper from compromising the secrecy of the ongoing transmissions. Trusted nodes allow scalable, secure communication, thus overcoming the restrictions imposed by distance-limited pairwise QKD schemes. A downside of the trusted node QKD is that its security guarantee is based on the assumption that \emph{all} intermediate nodes can be trusted. The assumption of having a dedicated quantum channel for each classical link in the network will be relaxed later in Section \ref{multilevel}, where we study flows with varying degrees of security requirements. 
%can be partially relaxed by utilizing the path diversity offered by our multi-path routing policy, described later in Section \ref{secD}.  

From the point of view of resource allocation, each link in a QKD network can be thought to be equipped with two distinct resources - (A) the physical link capacity and (B) the residual quantum keys available for encryption. While the first resource remains constant with time, the latter resource is time-varying and critically depends on the routing policy used. In order to achieve the maximum possible end-to-end throughput, the policy must utilize both the resources in an optimal fashion. Throughput-optimal policies for classical networks, such as Back Pressure \cite{tassiulas1990stability} or Universal Max-Weight \cite{sinha2017optimal}, optimize the routing policy with respect to the link capacities (A) only. The additional resource constraint, stemming from the availability of the residual quantum keys, is unique to the QKD Networks, which we address in this paper.  
 
%Unlike regular unicast-type flows where each packet has a unique destination node, in broadcast and multicast-type flows, a single packet needs to be delivered to multiple nodes simultaneously in the network. Broadcasting and multicasting are essential primitives in tactical military networks where a packet needs to be securely transmitted from a command and control center to multiple terminal nodes. These types of traffic are also common in emerging applications such as video-conferencing and live data-streaming. However, to the best of our knowledge, no capacity achieving routing policy is known for QKD networks that supports broadcast or multicast-type traffic.

\paragraph*{Related work} To achieve the network-layer capacity of a multi-hop network, one must use multi-path routing in an optimal fashion that is commensurate with the external packet arrival rates. In a seminal paper \cite{tassiulas1990stability}, Tassiulas and Ephremides proposed the celebrated \emph{Back-pressure} algorithm, which was shown to be throughput-optimal for unicast traffic. Numerous extensions and enhancements to the basic Back-pressure scheme have been proposed in the literature for the last thirty years \cite{neely2010stochastic, bp-survey1}. Unlike the regular unicast-type flows where each packet has a unique destination node, in broadcast and multicast-type flows, a single packet needs to be delivered to multiple nodes simultaneously in the network. The Back-pressure policy works for unicast flows only and it does not support broadcast or multicast traffic \cite{sarkar2002framework}. Using the Back-pressure algorithm as a building block, the paper \cite{zhou2019security} proposes a quantum key management and unicast routing policy to maximize the utility of a QKD network.  
In addition to being limited to unicast flows only, a major technical limitation of the scheme of \cite{zhou2019security} is that, in order to stabilize the data queues, the authors place an artificial constraint on the number of keys a node can have at its disposal \cite[Lemma 1]{zhou2019security}. However, unlike the data packets in transit, the abundance of quantum keys is always desirable as they can be used to encrypt more data packets. Hence, the performance of the algorithm proposed in \cite{zhou2019security} could be sub-optimal. The paper \cite{qkd-net} proposes a heuristic shortest path-based routing policy for a classically fully-connected network in the trusted-node QKD setting and provides simulation results. The papers \cite{tanizawa2016routing, han2014novel} propose a similar routing policy for key relaying by identifying paths containing sufficiently many residual keys. However, to the best of our knowledge, the problem of designing a provably throughput-optimal routing and scheduling policy for generalized traffic flows in the QKD networks is still open.  

\paragraph*{Our contributions} Building upon our previous work on the Universal Max-Weight (UMW) policy \cite{sinha2017optimal}, in this paper, we propose \textsc{Tandem Queue Decomposition} (TQD) policy that securely supports any arrival rate within the interior of the secure capacity region of a QKD network in the trusted node setup. The TQD policy concurrently supports diverse types of network traffic including unicast, broadcast, and multicast. The proposed policy is fully online and does not need to know either the packet arrival rates or the quantum key generation rates. If the arrival rate vector lies outside the capacity region, usual admission control mechanisms, such as the one proposed in \cite{sinha2018network}, may be used in conjunction with the algorithm developed in this paper.

%In a quantum network with trusted nodes (repeaters) we can ensure secure transmission of data over larger distances \cite{elliott2004darpa, elliott2002building}. But as network traffic increases it is essential to have multipath routing and scheduling of packets to avoid overloading of links and improving latency. In this paper, we will be using quantum channel to share the keys and classical channel to share the message signal, as shown in fig \ref{linkin}. 

%We will attempt to replace the much celebrated Back-pressure algorithm \cite{tassiulas1990stability} used currently in QKD network \cite{zhou2019security}, with the Universal Max Weight (UMW) algorithm \cite{sinha2017optimal, sinha2019throughput, sinha2018network} to improve (i) efficiency (ii) broadcasting/multicasting capability (iii) better delay properties and (iv) reduction of state-complexity.

%Our policy, which we call \textsc{Tandem Queue Decomposition} (TQD), 
Technically, the \textsc{TQD} policy maintains a virtual network of queues, each of which is implemented as a vector of counters. The main ingredient of the TQD policy is a new queueing architecture consisting of \emph{two} virtual queues in tandem for each communication link in the network. The reader should compare this architecture with the original UMW architecture \cite{sinha2017optimal} that defines \emph{one} virtual queue per link. In the case of QKD networks, the second virtual queue is essential to account for the transmission constraint imposed by the availability of residual quantum keys. We refer the reader to section \ref{secD} for a detailed description of the construction of the virtual queues. The TQD policy employs the UMW policy on a transformed network containing twice as many edges as the original network. The route of each packet (\emph{e.g.,} a path or tree depending on whether the packet belongs to unicast, multicast, or broadcast flow) is chosen dynamically using ``weighted-shortest-path" computations on the transformed network.  
%We show that the proposed policy is throughput-optimal while guaranteeing the unconditional security of Quantum Key Distribution. 
%Unlike \cite{zhou2019security}, TQD does not limit the number of keys on a key bank and admits loop-free routing, leading to a better delay performance.

%Unlike UMW policy, we add a virtual node between every link to assist the decoupling of encryption and data transmission.

%Building upon our earlier work on the Universal Max-Weight algorithm \cite{sinha2017optimal}, in this paper, we propose an efficient universal routing policy (TQD) that does not limit the number of keys in a node yet achieves the full throughput region of the network. Our policy supports a wide range of traffic types, including unicast, broadcast, and multicast. Furthermore, since TQD admits loop-free routing, it offers a better delay performance than \cite{zhou2019security} for unicast flows. To achieve this, TQD uses the UMW policy on a transformed network with twice the number of edges compared to the original network. 

The rest of the paper is organized as follows: In section \ref{secB}, we describe the system model and formulate the problem precisely. In section \ref{secD}, we give a brief overview of the TQD policy and describe the dynamics of the virtual queues on which TQD is based. In section \ref{secE}, we show that the proposed policy achieves the entire secured throughput region for any arbitrary network having a wide class of traffic. An extension of the proposed policies to heterogeneous networks, which contains both encrypted and unencrypted traffic, has been discussed in section \ref{multilevel}.  
 In section \ref{sims}, we compare the performance of the TQD policy with a few other competing routing policies using a simulator that we built on top of the state-of-the-art OMNeT++ platform. Finally, we conclude the paper in section \ref{secI} with some directions for future investigations.

\section{System Model and Problem Formulation} \label{secB}
In this section, we describe a simplified model of trusted node QKD networks built with point-to-point overlay QKD links. Since we are primarily concerned with the network layer aspects of QKD networks, some physical layer issues have been abstracted away in this model. 
%In the following, we define the notion of the secure capacity region in this context.
 Extension of the basic model to more practical heterogeneous networks with multiple security levels are described later in Section \ref{multilevel}. 
\subsection{Network Model}
We consider a network with arbitrary topology, represented by a graph $\mathcal{G}(V,E)$, where $V$ denotes the set of nodes ($|V| = n$) and $E$ denotes the set of edges ($|E| = m$). The edges could be either directed or undirected. Time evolves in discrete slots. Each edge in the network encompasses two types of links - a classical link and a QKD link. The capacity of the classical link $e$ is $\gamma_e$, \emph{i.e.,} it can transmit $\gamma_e$ number of encrypted packets per slot. The QKD links are used for symmetric quantum key agreement between the nodal end-points and \emph{not} for the actual data transfer, which takes place over the physical links. Although the network topology is assumed to be static, our proposed policy works even for time-varying networks. Furthermore, all our results can be straightforwardly generalized to networks with scheduling constraints (\emph{e.g.,} wireless networks).

%Here we introduce a new node within the transmitting node to assist decoupling the two steps involved in sending a data packet over an edge: (i) Encryption with Quantum keys (ii) Physical transmission of encrypted keys. This is a core aspect to our model. Thus we effectively double the number of links used in analysis by introducing a new node between every pair of transmitter and receiver nodes. To visualise better we can simply say that node $a_1$ within $A$ is where the packets are stored initially, and once they are encrypted they move to node $a_2$. Thus visualising nodes $a_1$ and $a_2$ as different nodes with a link between them is a mere abstraction we use to solve the problem. 

%\begin{figure}[h]
%\centering\includegraphics[width=0.47\textwidth]{ModelTandem.png}
%\put(-234,48){$\scriptsize{\eta_e}$}
%\put(-175, 94){$\gamma_e$}
%\caption{Consider a data packet moving sequentially from Node $A$ to $B$ to $C$. The figure above represents the physical setup between the nodes and the figure below is the algorithmic abstraction. The link between two arbitrary trusted nodes $A$ and $B$ comprises of the classical data link for packet transfer, the quantum link for mutual key agreement, and an abstract link to intermediary node where encrypted data is stored before sending via the classical link. 
%}
% \label{linkin}
%\end{figure}
%

\subsection{Quantum Keys - Generation, Distribution, and Consumption}
%The keys generated by the quantum entanglement mechanism between two trusted nodes are shared via the overlay QKD channels. 
We assume that pairwise secret keys are continuously generated between each node pair connected by QKD links. For completeness, we briefly review polarization-based prepare and measure BB84 protocol in Appendix \ref{bb84}.
The generated keys are stored on \emph{key banks}, which are typically implemented with text files located at each node \cite{evans}. Note that the key banks are different from the data queues; while the data queues hold physical data packets, the key banks store private symmetric quantum keys, which are used for encrypting the data packets before each transmission (see Figure \ref{linkin1}). Due to the inherent randomness of the key generation process, noise, and possible eavesdropping activity on the quantum channel, the amount of secret quantum keys generated per slot varies randomly. Let $K_{e}(t)$ be the number of keys generated over the QKD link $e$ at the time slot $t$. In this paper, we assume that $\{K_e(t)\}_{t\geq 1}$ is an i.i.d. stochastic process with $\mathbb{E}(K_e(t))= \eta_e$ such that $0\leq K_e(t) \leq K_{\max}, \forall t, e$ for some finite constant $K_{\max}$. When QKD is used in conjunction with a standard symmetric cipher, such as AES-128, the length of the plaintext message that can be encrypted by reusing a given amount of key material depends on the configured key renewal rate of the cipher\footnote{As an example, the cipher module produced by Xilinx uses a $128$-bit key to encrypt $\sim 2 ~$Gbit of plaintext \cite{alleaume2014using}.}. To simplify the notations, we normalize the key generation unit so that one unit of key encodes precisely one data packet. Due to the technological and physical challenges arising from entanglement generation, quantum decoherence, and implementation non-idealities, the quantum key generation is usually the bottleneck for information transmission \cite{ekert1991quantum, nauerth2013air, mailloux2015modeling}. Since the abundance of encryption keys is always desirable, we do not impose any hard upper limit on the size of the key banks (c.f. \cite{zhou2019security}). 
%Thus, unlike \cite{zhou2019security}, we do not require the key banks to be stable. 
Our objective is to design a policy that stabilizes the data queues for any arrival rate within the secure capacity region defined below. 

% Let $K_{e}(t)$ be the number of private symmetric keys generated for the link $e$ in the time slot $t$, where $\mathbb{E}(K_e(t))= \eta_e$ and $K_e(t) \leq K_{\max}$ for some finite constant $K_{\max}$.
 % At a time slot $t$ we only encrypt up to a maximum of $K_e(t)$ packets in the policy. Even though there might be wastage of a few keys we achieve throughput optimality, and later we go on to show how introducing storage without changing the policy will result in much better delay performance. 

%The encrypted keys are moved onto another queue (represented by the sub-nodes with subscript 2 in Figure \ref{linkin}) where we assume no physical upper limit on how many data packets can move along the abstract link (as this is not an actual link but a process).
\subsection{Data Traffic Model}
We consider a generalized traffic model, where a data packet arriving at a source node $s$ can either have a single destination (\textsf{Unicast}), or multiple destinations (\textsf{Multicast}). A special case of \textsf{Multicast} traffic is \textsf{Broadcast}, where an incoming packet is required to be delivered to all nodes in the network. Formally, we categorize the incoming packets into multiple classes $\mathcal{C}$ depending on its source $s^{(c)}$ and the set of destination(s) $\mathcal{D}^{(c)}$.  
Packets belonging to the class $c$ are assumed to arrive at the source i.i.d. at every slot at the rate $\lambda^c$. In other words, if $A^{(c)}(t)$ denotes the number of external packets from class $c$ that arrives at the source $s^{(c)}$, we have $\mathbb{E}A^{(c)}(t) = \lambda^{(c)}, \forall c \in \mathcal{C}$. The joint arrival rate vector $\boldsymbol{\lambda}$ is obtained by concatenating the arrival rates of each class, \emph{i.e.,} $\boldsymbol{\lambda} = (\lambda_1, \lambda_2, \ldots, \lambda_{|\mathcal{C}|})$. 
We also assume that the total number of new packet arrivals to the entire network at any time slot is bounded by a finite constant $A_{\text{max}}$.

\subsection{Policy Space}
An admissible policy for this problem is responsible for the following operations - (1) selecting a route for each packet based on its traffic class and possibly duplicating the packet along its way as necessary (in the case of broadcast/multicast traffic), (2) encrypting the link traffic with the available quantum keys, and (3) forwarding the encrypted packets over the classical communication links. Note that a data packet can be forwarded over a link only if sufficiently many quantum keys are available for encryption. Otherwise, the packet must wait until the keys are generated. The set of all admissible policies is denoted by $\Pi$. 
%The set $\Pi$ also includes clairvoyant policies, which may use future packet arrival information.

We say that a policy $\pi \in \Pi$ \emph{securely} supports an arrival rate vector $\bm{\lambda}$ if under the action of the policy $\pi$, the destination node(s) of class $c$ receive(s) encrypted class $c$ packets at the rate $\lambda^{(c)}$, $\forall c \in C$. Formally, let $R^{(c)}(t)$ denote the total number of class $c$ packets commonly received by the destination node(s) $\mathcal{D}^{(c)} $ under the action of the policy $\pi$ up to time $t$. We now make the following definitions.

\begin{defi} 
[\textbf{Policy Securely Supporting an Arrival Rate Vector} $\boldsymbol{\lambda}$] \label{uiy}
A policy $\pi \in \Pi$ is said to securely support an arrival rate vector $\boldsymbol{\lambda}$ if 
\begin{align*}
	\liminf_{t \to \infty}\frac{R^{(c)}(t)}{t} = \lambda^{(c)}, \quad \forall c \in \mathcal{C},\quad \text{w.p.}~1
\end{align*}
\end{defi}

%\subsection{Network-Layer Capacity Region}
\begin{defi}
[\textbf{Stability Region of a Policy}] \label{stability}	
The stability region $\Lambda_\pi(\mathcal{G}, \bm{\eta}, \bm{\gamma})$ of an admissible policy $\pi$ is defined to be the set
of all arrival rate vectors securely supported by the policy $\pi,$ \emph{i.e.,}
\begin{align*}
	\boldsymbol{\Lambda_{\pi}}(\mathcal{G},\bm{\eta}, \bm{\gamma}) \stackrel{\textrm{(\emph{def})}}{=} \{ \boldsymbol{\lambda} \in \mathbb{R}_+^{|\mathcal{C}|} : \pi  \textrm{ \emph{securely supports} }  \boldsymbol{\lambda}\},
\end{align*}
\end{defi}
where $\mathbb{R}_+$ denotes the set of all non-negative numbers. 
%The stability region $\Lambda_\pi$ of an admissible policy $\pi$ is defined to be the set
%of all arrival rate vectors securely supportable by the policy $\pi$.
In the above definition, we have made the dependence of the stability region with the network topology ($\mathcal{G}$), key-generation rates ($\bm{\eta}$), and the link capacities ($\bm{\gamma}$) explicit. 
The secure capacity region $\bm{\Lambda}(\mathcal{G}, \bm{\eta}, \bm{\gamma})$ is defined to be the set of all arrival rate vectors supported by an admissible policy. Formally, 
\begin{defi} 
[\textbf{Secure Capacity Region of a Network}]\label{net-cap-def}
The secure capacity region $\boldsymbol{\Lambda}(\mathcal{G},\bm{\eta}, \bm{\gamma})$ of a network is defined to be the set of all supportable rates, i.e.,
\begin{align*}
	\boldsymbol{\Lambda}(\mathcal{G},\bm{\eta}, \bm{\gamma}) = \bigcup_{\pi \in \Pi} \boldsymbol{\Lambda_{\pi}}(\mathcal{G},\bm{\eta}, \bm{\gamma}).
\end{align*}
\end{defi}
Finally, we define the notion of a \emph{secure} throughput-optimal policy, which generalizes the notion of throughput-optimal policies given in \cite{tassiulas1990stability}. 
%Note that, two different rate vectors in the secured capacity region could be achieved by two different admissible policies. 
%A policy $\pi_1$ dominates another policy $\pi_2$ if $\Lambda_{\pi_{1}} \subseteq  \Lambda_{\pi_{2}}$ \cite{tassiulas1990stability}.

%An optimal policy is the one which dominates any other policy in $\Pi$, should have stability region that is a super-set of the
%stability region of any other policy in $\Pi$. Therefore, it
%should have stability region equal to $\boldsymbol{\Lambda}$. Such a policy is called a maximum throughput policy \cite{tassiulas1990stability}. Formally we define:

%The set $\boldsymbol{\Lambda}(\mathcal{G},\mathcal{C})$ is convex.

\begin{defi}
[\textbf{Secure Throughput-Optimal Policy}] A \textit{secure throughput-optimal policy} is an admissible policy $ \pi^* \in \Pi $, that supports any arrival rate $\boldsymbol{\lambda}$ in the interior of the secure capacity region $\boldsymbol{\Lambda}(\mathcal{G},\bm{\eta}, \bm{\gamma})$. 
\end{defi}
From the above definition, it is unclear whether a secure throughput-optimal policy exists as two different rate vectors in the secure capacity region might not be achieved by the same admissible policy. One of the major contributions of the paper is to show that a secure throughput-optimal policy exists, and it can be efficiently implemented.  

\section{Characterization of the Secure Capacity Region}
Let $\mathcal{G}_{\boldsymbol{\omega}}$ be the  capacitated version of the given network where the capacity $\omega_e$ for the edge $e$ is defined as follows:
\begin{align*}
    {\omega}_{e} &= \min(\gamma_e,\eta_e), \quad \quad \forall e \in E.
\end{align*}
In Theorem \ref{cap_ch} below, we show that the cap acity region of the network is given by the set of all feasible generalized multi-commodity flow vectors in the capacitated graph $\mathcal{G}_{\bm{\omega}}.$
One direction of this result is quite intuitive; the long-term rate of encrypted packet flow over an edge $e$ is limited by the quantum key generation rates and the capacity of the communication link $e$.  
Consider an arrival rate vector $\boldsymbol{\lambda} \in \boldsymbol{\Lambda}(\mathcal{G},\bm{\eta}, \bm{\gamma})$. By definition, there exists an admissible policy $\pi \in \Pi$ that supports the arrival rate $\boldsymbol{\lambda}$. Upon taking a long-term time-average over the actions of the policy $\pi$, it is evident that we can obtain a \textit{randomized} flow decomposition on $\mathcal{G}_{\omega}$ such that none of its edges are overloaded. In other words, for every $\boldsymbol{\lambda} \in \boldsymbol{\Lambda}(\mathcal{G},\bm{\gamma}, \bm{\eta})$, there exist a non-negative scalar $\lambda^{(c)}_i$, associated with the $i$\textsuperscript{th} admissible route $T^{(c)}_i \in \mathcal{T}^{(c)}, \forall i, c,$ such that
\begin{align}
    \lambda^{(c)} &= \sum_{i:T^{(c)}_i \in \mathcal{T}^{(c)}}\lambda_i^{(c)}, \label{wde}\\
    \lambda_e &\stackrel{\textrm{(def.)}}{=} \sum_{\substack{(i,c): e\in T_i^{(c)},\\
   T_i^{(c)} \in \mathcal{T}^{(c)}}}  \lambda_i^{(c)}
   \leq \omega_e, \quad \forall e \in E. \label{rdde}
\end{align}
 Eqn. \eqref{wde} shows that there exists such a valid flow decomposition across the routes. The inequality in \eqref{rdde} states that no edge in $\mathcal{G}_{\bm{\omega}}$ is overloaded. To formally state our result, we need the following definition of the feasible flow region $\overline{\boldsymbol{\Lambda}}_{\bm{\omega}}$ of $\mathcal{G}_{\bm{\omega}}$.
 %, i.e. arrival rate of packets to any edge $e$ under policy $\pi$ will be at most the physical capacity $\gamma_e$ of the edge or the quantum key generation rate $\eta_e$ of edge $e$, whichever is lesser. 
\begin{defi} \label{deftr}
The set $\overline{\boldsymbol{\Lambda}}_{\bm{\omega}}$ is defined as the set of all
arrival vectors $\boldsymbol{\lambda} \in \mathbb{R}^{|\mathcal{C}|}_+$ for which there exists a non-negative flow decomposition $\{\lambda_i^{(c)}, \forall i, c\}$ such that the inequalities \eqref{wde} and \eqref{rdde} are satisfied. 
\end{defi}
Let $\textrm{int}(\cdot)$ denote the interior of a subset of an $n$-dimensional Euclidean space. The following theorem characterizes the secure capacity region of a network.
%For any subset $S$ of the $k$-dimensional Euclidean space, let $\textrm{int}(S)$ denote the interior of the set $S$ respectively \cite{rudin1964principles}. 
%We now state the main result of this paper:
%Using Definition \eqref{deftr}, we have the following theorem:
\begin{theo} [Characterization of the Secure Capacity region] \label{cap_ch}
The Secure Capacity region $\boldsymbol{\Lambda}(\mathcal{G},\bm{\eta}, \bm{\gamma})$ is identical to the set $\overline{\boldsymbol{\Lambda}}_{\bm{\omega}}$, up to its boundary, i.e., the following two set inclusions hold:
\begin{enumerate}
\item  \textbf{\emph{[Converse]}} $\bm{\Lambda} \subseteq \overline{\bm{\Lambda}}_{\bm{\omega}}$.
\item \textbf{\emph{[Achievability]}} $\textrm{\emph{int}}(\overline{\boldsymbol{\Lambda}}_{\bm{\omega}}) \subseteq \bm{\Lambda}$ and there exists an admissible policy which achieves any rate within the set $\textrm{\emph{int}}(\overline{\boldsymbol{\Lambda}}_{\bm{\omega}})$.
\end{enumerate}
\end{theo}
The proof of the converse is given in Appendix \ref{app1}.
%Appendix A of the extended version of this paper \cite{quantnet2020Vishnu}. 
The achievability result is more interesting from a policy design point of view. We establish the achievability result by designing an efficient policy, called \textsc{TQD},  that supports any rate within the set $\textrm{int}(\overline{\boldsymbol{\Lambda}})$. 

\section{Designing a Secure Throughput-Optimal Policy} \label{secD}

%\section{Secure Capacity Region} \label{capacity}
A fundamental difference between the two major resources in a QKD network, namely the link capacity and the residual quantum keys, is that, unlike the former, the latter can be stored for future use. Thus, the number of quantum keys available for encryption over any physical link depends critically upon the packet routing policy employed in the network. More explicitly, if a policy routes the majority of the packets over a small subset of physical links, the quantum keys corresponding to that set of links will get exhausted quickly. This observation should be contrasted with the physical link capacities which are independent of the routing policy. The dependence of the amount of residual quantum keys on the routing policy makes the system non-memoryless. As a result, analyzing and controlling the QKD network, and establishing the optimality of a particular policy becomes significantly more challenging than the classical networks.

\paragraph*{A counterexample} To illustrate the non-triviality of the problem, consider a simple candidate policy $\pi^{\textrm{single queue}}$ using a \emph{single queue} per link that (1) encrypts packets with only freshly generated quantum keys for the overlay QKD links, (2) forwards the encrypted packets \emph{immediately} over the physical links using some routing policy, and then (3) discards any unused quantum keys at the end of each slot. Hence, the policy $\pi^{\textrm{single queue}}$ does not store quantum keys for future use and, consequently, is much easier to analyze, thanks to the i.i.d. nature of the key generation process. Unfortunately, as we show in the example below, the policy  $\pi^{\textrm{single queue}}$ is \emph{not} throughput-optimal. Nevertheless, we will soon see that this example suggests a new secure throughput-optimal architecture for any arbitrary QKD network. 

\begin{figure}
\centering
 \includegraphics[height=0.08\textwidth]{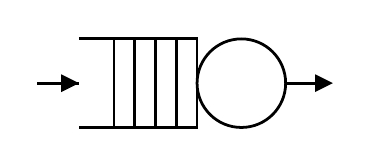}	
 \put(-162,19){\textrm{\small{Unencrypted packets}}}
 \put(-33, 2){$\uparrow$}
 \put(-50, -8){\textrm{\small{Quantum keys~}}$K(t)$}
 \put(-75, 40){\small{Service rate $=\min(\gamma, K(t))$}}
 \put(-3,19){\textrm{\small{Encrypted packets}}}
 \caption{\footnotesize{Depicting a single QKD link under the policy $\pi^{\textrm{single queue}}$}}
 \label{single-qkd-link}
\end{figure}
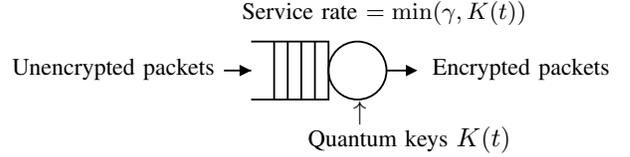

\paragraph*{Sub-optimality of $\pi^{\textrm{single queue}}$} Consider a point-to-point QKD setup where the physical channel has capacity equal to $\gamma$ and the overlay QKD channel has key generation rate $\eta$ (please refer to Figure \ref{single-qkd-link}). Let the random variable $K(t)$ denote the number of quantum keys generated by the QKD link at time slot $t$ s.t. $\mathbb{E}K(t)=\eta$. As in our system model, we assume that the sequence of r.v.s $\{K(t)\}_{t\geq 1}$ is i.i.d. across slots. Since the policy $\pi^{\textrm{single queue}}$ discards any unused keys at the end of every slot, the overall service process from the link is i.i.d. with mean value $\mu^{\textrm{single queue}}= \mathbb{E}(\min(\gamma, K(t)).$ Using Jensen's inequality, we can immediately conclude that
\begin{eqnarray} \label{jensen}
	\mu^{\textrm{single queue}}&=&\mathbb{E}(\min(\gamma, K(t)) \nonumber \\
	&\stackrel{(a)}{\leq}& \min(\gamma, \mathbb{E}K(t)) = \min(\gamma, \eta).
\end{eqnarray}
Furthermore, the inequality (a) in Eqn.\ \eqref{jensen} could be strict. For example, consider the case $\gamma=1$ and $K(t) \sim \textrm{Poisson}(\eta)$ with $\eta = \nicefrac{1}{2}.$ In this case, we have:
\begin{eqnarray*}
\mathbb{E}(\min(\gamma, K(t)))= 1-e^{-0.5} \approx 0.393 < 0.5=\min(\gamma, \eta).	
\end{eqnarray*}
From basic queueing theory \cite{wolff1989stochastic}, the maximum rate $\lambda^{\textrm{single queue}}_*$ achievable by the policy $\pi^{\textrm{single queue}}$ is given by the service rate, i.e., $\lambda^{\textrm{single queue}}_* \approx 0.393$. On the other hand, by using the TQD architecture described below, we will show that a rate of $0.5$ is achievable for the above simple single link setting. This shows that the policy $\pi^{\textrm{single queue}}$ is not capacity-achieving.

In the following, we describe an efficient routing and key management policy that achieves the entire secure capacity region of any given network. As discussed above, the key availability constraint makes this problem more challenging than the vanilla universal network flow problem considered in  \cite{sinha2017optimal}. We solve this problem using a novel Tandem Queue Decomposition (TQD) framework that reduces the problem to an instance of the universal network flow problem without the key availability constraint.  
\subsection{The Tandem Queue Decomposition Architecture (TQD)}
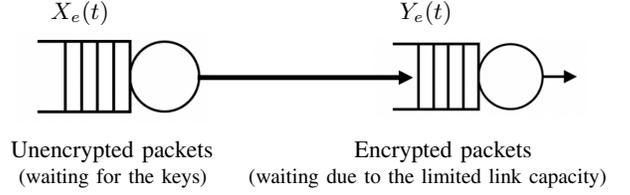
\begin{figure}
          \centering
       \includegraphics[height=0.08\textwidth]{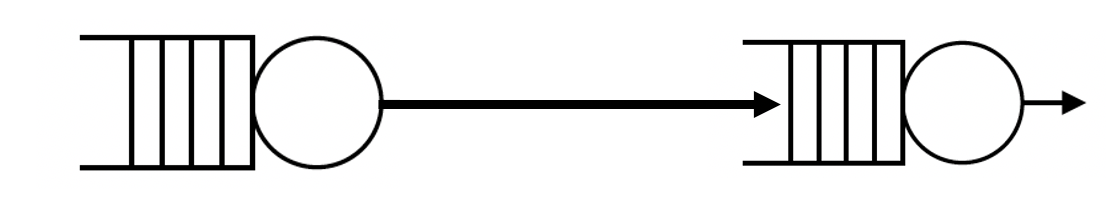}
       \put(-220,-10){\textrm{\small{\textcolor{black}{Unencrypted packets}}}}
       \put(-217, -20){\textrm{\footnotesize{(waiting for the keys)}} } 
       \put(-205, 42){\textrm{\small{$X_e(t)$}} }
       \put(-73, 42){\textrm{\small{$Y_e(t)$}} }
       \put(-90,-10){\textrm{\small{\textcolor{black}{Encrypted packets}}}}
       \put(-130, -20){\textrm{\footnotesize{(waiting due to the limited link capacity)}}}
          \caption{\small{TQD architecture for a single link $e$ consisting of a classical and an overlay point-to-point QKD link}}
                    \label{fig:tqd1}
 \end{figure}

To enforce the constraint that only encrypted packets can be transmitted over the physical links, we conceptually construct a \emph{transformed network} where every edge is split into two edges in tandem, each containing one queue. The first queue $X$, which is internal to the nodes, holds unencrypted packets waiting for the residual quantum keys. The second queue $Y$ holds encrypted packets waiting to be transmitted over the physical links. See Figure \ref{fig:tqd1} for the TQD architecture corresponding to a single link. We illustrate the construction of the transformed network via the following  example.  

\begin{figure}[h]
\centering\includegraphics[width=0.47\textwidth]{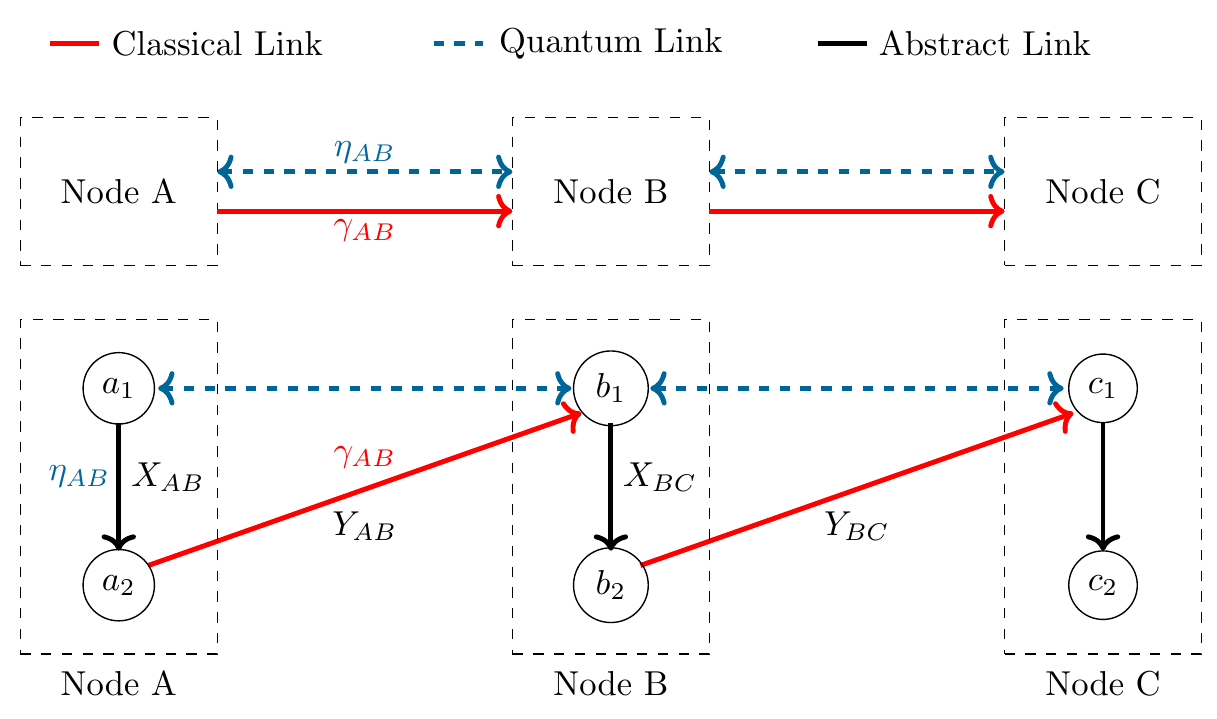}
%\put(-240,48){$\scriptsize{\textcolor{blue}{\eta_{AB}}}$}
\put(-252, 102){\small{\textrm{(A)}}}
\put(-252, 45){\small{\textrm{(B) }}}
%\put(-305,45){\small{\textrm{Network}}}
\put(-231, 75){\tiny{\textrm{Unencrypted}}}
\put(-229, 15){\tiny{\textrm{Encrypted}}}
%\put(-175, 94){$\textcolor{blue}{\tiny{\gamma_{AB}}}$}
\caption{\footnotesize{Consider a data packet moving from Node $A$ to Node $C$ via Node $B$. Part (A) of the schematic shows the physical connections between the nodes, and part (B) depicts its algorithmic abstraction after the Tandem Queue Decomposition procedure. The link between two trusted nodes $A$ and $B$ consists of the classical data link for packet transfer, the quantum link for mutual key agreement, and an abstract link to an intermediate node where the encrypted data is stored before sending it via the communication link. }
}
 \label{linkin}
\end{figure}
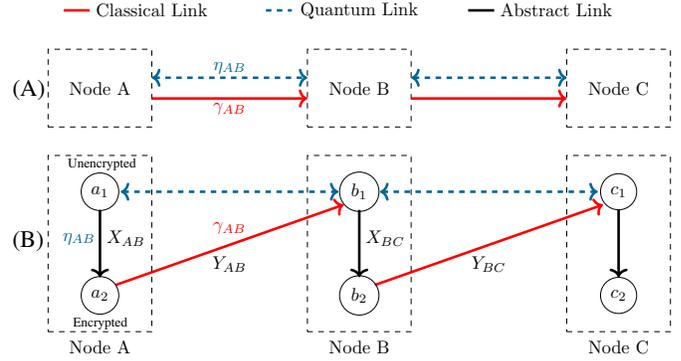

\paragraph*{Example} Consider an edge $(A,B)$ that connects node $A$ to node $B$ as shown in Figure \ref{linkin}. Assume that the quantum keys are generated over the link $(A,B)$ at the rate $\eta_{AB}$ and the corresponding physical communication link can transmit packets at the rate of $\gamma_{AB}$ packets per second. In the transformed network, we replace the edge $(A,B)$ by introducing two internal nodes $a_1$ and $a_2$, an \emph{internal} edge $(a_1, a_2)$ connecting them, and an \emph{external} edge $(a_2, b_1)$ as shown in Figure \ref{linkin}. A queue is associated with each of the newly introduced edges. The queue $X_{AB}$, corresponding to the internal edge $(a_1, a_2)$, holds the set of \emph{unencrypted} packets that are waiting for the quantum keys to become available. The queue $Y_{AB}$, corresponding to the external edge $(a_2, b_1)$, holds the set of \emph{encrypted} packets that are waiting to cross the physical link $AB$. Similar decompositions are performed for each link in the network. 
Since the above transformation does not alter the capacity region of the network, it is sufficient to design a throughput-optimal policy for the transformed network. In the following, we use the Universal Max-Weight policy \cite{sinha2017optimal} on the transformed network for accomplishing this goal. 
%Unless stated otherwise, in the rest of the paper, by the term "network", we refer to the TQD transformed network only.

%\cmt{draw a figure}
%\begin{figure}[h]
%\centering\includegraphics[width=0.47\textwidth]{ModelTandem.png}
%\put(-240,48){$\scriptsize{\textcolor{blue}{\eta_{AB}}}$}
%\put(-235, 76){\tiny{\textrm{Unencrypted}}}
%\put(-233, 16){\tiny{\textrm{Encrypted}}}
%\put(-175, 94){$\textcolor{blue}{\tiny{\gamma_{AB}}}$}
%\caption{Consider a data packet moving sequentially from Node $A$ to Node $B$ to Node $C$. The figure above represents the physical setup between the nodes and the figure below is its algorithmic abstraction after the decomposition is done by the TQD process. The link between two arbitrary trusted nodes $A$ and $B$ comprises of the classical data link for packet transfer, the quantum link for mutual key agreement, and an abstract link to intermediary node where encrypted data is stored before sending via the classical link. 
%}
% \label{linkin}
%\end{figure}

\subsection{Precedence Constraints and the Virtual Queueing Process}
\begin{figure}[h]
\centering
\includegraphics[width=0.35\textwidth]{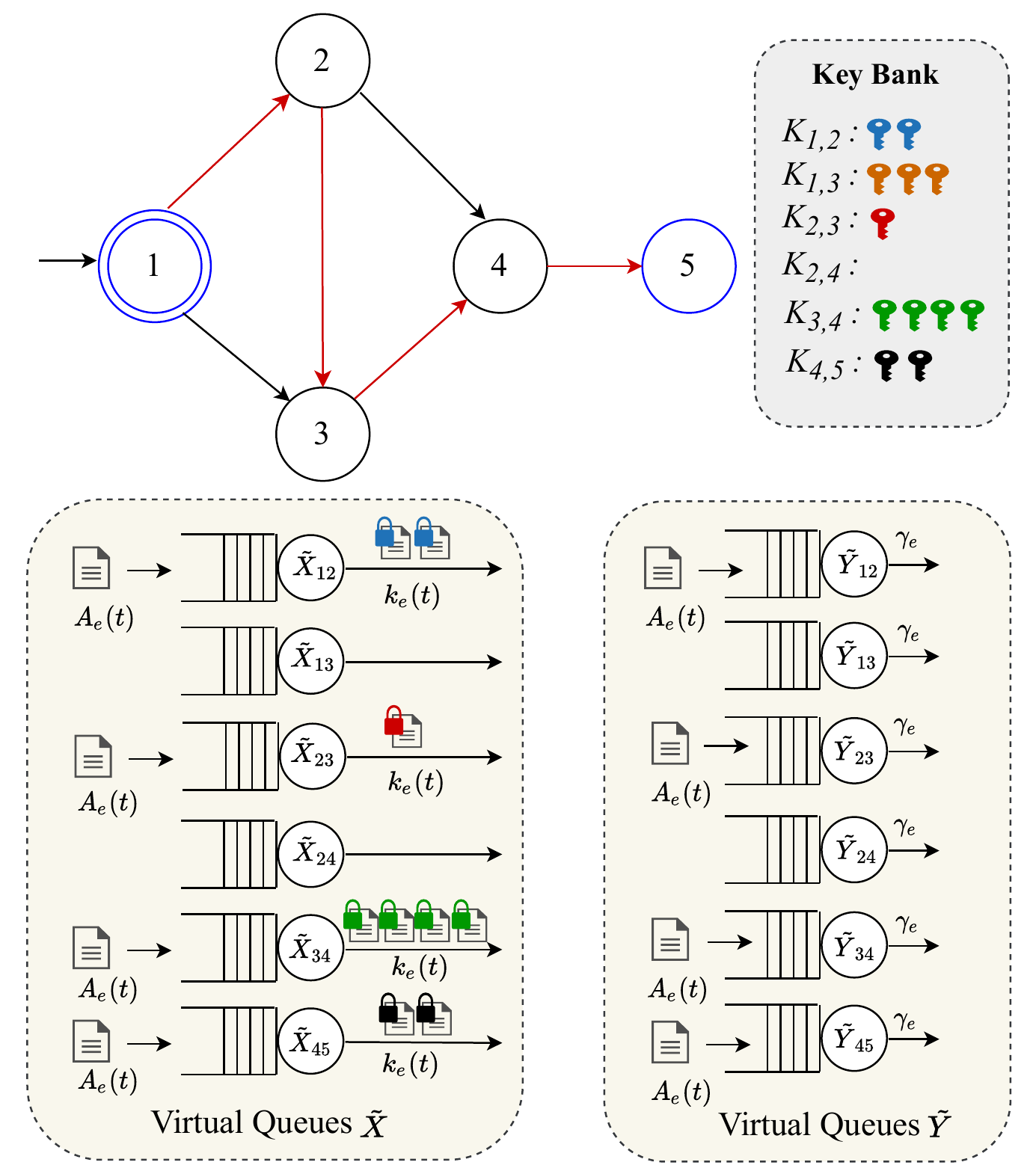}
\caption{\footnotesize{Illustration of the virtual queue dynamics for the five-node network $\mathcal{G}$. When a packet arrives at source 1 with destination 5, and given the assigned route $ \mathcal{T}_p =
\{\{1, 2\}, \{2, 3\}, \{3, 5\}\} $, the following queue updatings occur: 
The packet is counted simultaneously as an arrival to the virtual data queues $\tilde{X}_{12}, \tilde{X}_{23}, \tilde{X}_{35}, \tilde{Y}_{12}, \tilde{Y}_{23}, \tilde{Y}_{35}$ \textit{at the same slot}. The physical packet reaches these edges only at a subsequent time slot.}
}
\label{vq-fig}
\end{figure}

Due to the precedence constraints, a packet, which is being routed along the route $T = e_1-e_2-...-e_n,$ reaches the $j$\textsuperscript{th} link $e_j$ only after crossing the previous links on its path. Hence, the arrival process to a downstream queue depends on the state of the upstream queues in a complex fashion. As a result, directly analyzing and designing a stabilizing control policy for the real queueing system becomes challenging. To address this difficulty, similar to the \textsf{UMW} policy, we first relax the precedence constraints to obtain a single-hop \emph{virtual network}, which will be used for dynamically routing the incoming packets \cite{sinha2017optimal}. Towards this end, we define a $2m$-dimensional parallel virtual queueing process $\tilde{\bm{Q}}(t) := \{ \tilde{\bm{X}}(t),\tilde{\bm{Y}}(t) \},$ as shown in Figure \ref{vq-fig}. In this construction, we associate one virtual queue to each edge in the transformed network. Hence, the process $\tilde{\bm{X}}(t) = (\tilde{X}_e(t), e\in E)$ corresponds to the virtual queues holding the unencrypted packets waiting for the keys, and the process $\tilde{\bm{Y}}(t)=(\tilde{Y}_e(t), e\in E)$ corresponds to the virtual queues holding the encrypted packets waiting to be transmitted over the communication links in the virtual network. We emphasize that the virtual queues, which are just a set of numbers (state variables), follow simplified queueing dynamics without the precedence constraints, as detailed below.
%
%\begin{tcolorbox}
%\textbf{Note:} In the physical system, there are two tandem queues, with $X_e$ followed by $Y_e$ for each edge $e \in E$ as shown in Figure \ref{linkin}. The corresponding virtual queues are denoted by $\tilde{X}_e$ and $\tilde{Y}_e$. The packets in $X_e$ at slot $t$ are encrypted with the available quantum keys in the key bank and then forwarded to the physical queue $Y_e$. From $Y_e$, the packets get transmitted to the next link $e'$ and get queued at $X_{e'}$.
%\end{tcolorbox}
%

\paragraph*{Operation of the Virtual Queues} For each class $c$ packet, $c \in \mathcal{C},$ the TQD policy first decides a suitable route $T^{(c)}(t) \in \mathcal{T}^{(c)}$ immediately upon the packet's arrival. Let us denote the set of links on its prescribed route $T^{(c)}(t)$ by $\{e_i | i = 1,2, \ldots, k\}$. Each incoming packet induces a virtual packet arrival simultaneously at each of the virtual queues on its path, \emph{i.e.,} $\{\tilde{X}_{e_i} | i = 1,2, \ldots, k\}$ and $\{\tilde{Y}_{e_i} | i = 1,2, \ldots, k\}$. Unlike the physical system, which is limited by the precedence constraints, any packet present in the virtual queues is eligible for service immediately upon its arrival. Thus the number of packet arrivals $A_e^{\pi}(t)$ to both the virtual queues $\tilde{Q}_e = \{ \tilde{X}_e,\tilde{Y}_e\} $ at time $t$ under the action of a policy $\pi$ can be expressed as:
 \begin{align}
     A_e^{\pi}(t) = \sum_{c \in \mathcal{C}}A^{(c)}(t)\mathds{1}\big( e \in T^{(c)}(t)\big), \quad \forall e \in E.  \label{yhnb}
 \end{align}
 %The symmetric quantum key pairs generated at slot $t$ are stored into the key banks on both sides of the edge $e$.
  The unused quantum keys in the current slot are stored for the future use. Let $\kappa_e(t)$ denote the total number of keys available for encrypting the packets crossing the edge $e$ at slot $t$ and $k_e(t)$ denote the number of residual keys in the key bank available from the previous rounds. Hence, $\kappa_e(t) = K_e(t)+ k_e(t),$ where we recall that $K_e(t)$ is the number of \emph{new} quantum keys generated by the QKD link $e$ at slot $t$. 
 Note that the key availability process $\{\bm{\kappa}(t)\}_{t\geq 1}$ is heavily dependent on the routing policy used. Putting everything together, the one-step evolution of the virtual queue processes is given by the following two Lindley recursions:
 \begin{align} \label{lindley1}
 \tilde{X}_e(t+1) &= \big(\tilde{X}_e(t) + A^{\pi}_e(t) - \kappa_e(t)\big)^{+}, \quad \forall e \in E\\
 \tilde{Y}_{e}(t+1) &= \big(\tilde{Y}_{e}(t) + A_e^{\pi}(t)- \gamma_e\big)^{+}, \quad \forall e \in E. \label{lindley2}
 \end{align}

 %\section{TQD Algorithm } \label{secF}
 With the above description of the queueing architecture in place, we now present the Tandem Queue Decomposition policy in Algorithm \ref{tqd-algo}. The derivation of the policy and the proof of its throughput-optimality are given in the following section. The following salient features of the TQD policy are noteworthy.  
\begin{algorithm}
\begin{algorithmic}[1]
\REQUIRE Graph $\mathcal{G}(V,E)$, Virtual Queue lengths $\{\tilde{X}_e(t), e \in E \}$ and $\{\tilde{Y}_e(t), e \in E \}$ at the slot $t$
  \item[1:] \textbf{(Edge-Weight Assignment) } Assign each edge of the original 
graph $e \in E$ a weight $W_e(t)$ equal to $\tilde{X}_e(t)+\tilde{Y}_e(t)$, i.e., 
$$\boldsymbol{W}(t) \leftarrow\boldsymbol{ \tilde{X}}(t)+\boldsymbol{ \tilde{Y}}(t).$$
  \item[2:] \textbf{(Route Assignment)} For all incoming packets, compute a Minimum-Weight Route (according to the class of the packet) in the weighted graph $\mathcal{G}(V,E)$.
  \item[3:] \textbf{(Key Generation)} Generate symmetric private keys for every edge $e$ via QKD and store the keys in the key banks.
  \item[4:] \textbf{(Encryption)} Encrypt the data packets waiting in physical queue $X_e$ with the available keys in the key bank and internally transfer the encrypted packets to the downstream queue $Y_e$ for every edge $e \in E$.
  \item[5:] \textbf{(Packet Forwarding)} Forward the encrypted physical packets from the queue $Y_e$ to the queue $X_{e'}$ for every edge $e$ according to some packet scheduling policy (ENTO \cite{gamarnik1999stability}, FIFO etc). Here $e'$ is the next edge in the assigned route of a packet.
  \item[6:] \textbf{(Decryption)} Decrypt the data packets received at physical queue $X_e$ for every edge $e$ using the symmetric key generated earlier via the QKD process. 
  \item[7:] \textbf{(Updating the Virtual Queues)} Update the virtual queues assuming a precedence-relaxed system, i.e.,
  $$ \tilde{X}_e(t+1) \leftarrow \big(\tilde{X}_e(t) + A^{\pi}_e(t) - \kappa_e(t)\big)^{+},\quad \forall e \in E$$
  $$ \tilde{Y}_e(t+1) \leftarrow \big(\tilde{Y}_e(t) + A^{\pi}_e(t) - \gamma_e \big)^{+},\quad \forall e \in E.$$
\end{algorithmic}
\caption{Tandem Queue Decomposition (TQD) policy at slot $t$}
\label{tqd-algo}
\end{algorithm}

\begin{enumerate}
\item The routing policy is online as it is oblivious to the arrival rates $\bm{\lambda},$ the key generation rates $\bm{\eta}$, the physical capacities of the links $\bm{\gamma},$ and the physical queue lengths $(\bm{X}_e(t), \bm{Y}_t(t)), t\geq 1.$ 
\item The shortest path computations depend on the virtual queue lengths through the sum of the encrypted and unencrypted queues in each link and not on the individual virtual queue lengths.
\end{enumerate}

In the next section, we show that the proposed policy stabilizes both the virtual and the physical queues for all arrival rates within the interior of the secure capacity region $\bm{\Lambda}(\mathcal{G}, \bm{\eta}, \bm{\gamma})$. 
 
\section{Derivation of the TQD Policy and its Stability Properties} \label{secE}
%\paragraph{Stabilizing the Virtual Queues}
Due to the additional transmission constraints arising from the instantaneous availability of quantum keys in the key banks, the derivation and the proof of strong stability of the TQD policy differ significantly from that of the UMW policy \cite{sinha2017optimal}. To derive a stabilizing policy for the virtual network, we consider the following quadratic Lyapunov function $L(\tilde{Q}(t))$, defined in terms of the virtual queue lengths of both unencrypted and encrypted packets: 
\begin{eqnarray} \label{lyap}
L(\tilde{Q}(t)) = \sum_{e \in E} \big(\tilde{X}_{e} ^2(t)+ \tilde{Y}_{e} ^2(t)\big).
\end{eqnarray}
The Lyapunov function \eqref{lyap} defined above should be contrasted with the Lyapunov function used in Eqn.\ (9) of \cite{zhou2019security}, which includes the number of residual keys in the second term of the quadratic. Thus, any drift-minimizing policy of their Lyapunov function implicitly stabilizes the number of quantum keys in the key banks as well, which might result in sub-optimal throughput. 
From the one-step dynamics given in Eqns.\ \eqref{lindley1} and  \eqref{lindley2}, we have the following bounds:
\begin{align*}
         \tilde{X}_{e}(t+1)^2 &\leq \tilde{X}_{e}(t)^2 + {A_e^{\pi}}^2(t) + \kappa_e(t)^2 + 2\tilde{X}_{e} (t) A_e ^{\pi}(t) \\ 
        & \quad - 2\tilde{X}_{e}(t)\kappa_e(t) - 2A_e ^{\pi}(t)\kappa_e(t),\\
          \tilde{Y}_{e}(t+1)^2 &\leq \tilde{Y}_{e}(t)^2 + {A_e^{\pi}}^2(t) + \gamma_e^2 + 2\tilde{Y}_{e} (t) A_e ^{\pi}(t) \\ 
        & \quad - 2\tilde{Y}_{e}(t)\gamma_e - 2A_e ^{\pi}(t)\gamma_e.
\end{align*}
%\cmt{use the fact that $\tilde{X}_e(t) K_e(t)=0.$}
Since $\tilde{X}_e(t) \geq 0, \kappa_e(t) \geq 0$, $A_e^{\pi}(t) \geq 0$ and $\gamma_e \geq 0$, we can write:
\begin{align}
    \tilde{X}_{e}(t+1)^2 -  \tilde{X}_{e}(t)^2 &\leq {A_e^{\pi}}^2(t) + \kappa_e(t)^2  \nonumber \\
    & \quad + 2\tilde{X}_{e}(t)A_e ^{\pi}(t) - 2\tilde{X}_{e}(t)\kappa_e(t),\label{Xq}\\
    \tilde{Y}_{e}(t+1)^2 -  \tilde{Y}_{e}(t)^2 &\leq {A_e^{\pi}}^2(t) + \gamma_e^2 \nonumber \\
    & \quad + 2\tilde{Y}_{e}(t)A_e ^{\pi}(t) - 2\tilde{Y}_{e}(t)\gamma_e \label{Yq}.
\end{align}
%Let $\kappa_e(t)$ be the number of unused quantum keys remaining in the key-bank from the previous round and $k_e(t)$ be the random number of new quantum keys generated at slot $t$. 
%Clearly, $\kappa^\pi_e(t) \geq K(t).$
Next, we observe that we always have $\tilde{X}_e(t) k_e(t) =0.$ This equation can be understood as follows. Since all currently available keys are used for encryption, if there are packets in the queue waiting to be encrypted (\emph{i.e.,}  if $\tilde{X}_e(t) >0$), there cannot be any residual keys from the previous round (\emph{i.e.,} $k_e(t)=0$.) Since $\kappa_e(t) = k_e(t)+ K_e(t)$, for $\tilde{X}_e(t)>0,$ we can rewrite the inequality \eqref{Xq} as:
\begin{align}
    \tilde{X}_{e}^2(t+1) -  \tilde{X}_{e}^2(t) &\leq {A_e^{\pi}}^2(t) + K_e^2(t)  \nonumber \\
    & \quad + 2\tilde{X}_{e}(t)A_e ^{\pi}(t) - 2\tilde{X}_{e}(t)K_e(t).\label{Xq1}   
    \end{align} 
    On the other hand, if $\tilde{X}_e(t)=0,$ trivially we have $\tilde{X}_e(t+1) \leq A_e^{\pi}(t).$ Thus, even in this case, the bound in Eqn.\ \eqref{Xq1} continues to hold. 
Combining Eqns.\ \eqref{Yq} and \eqref{Xq1} with the fact that $\mathbb{E}(K_e(t)|\tilde{\bm{Q}}(t))= \mathbb{E}K_e(t)= \eta_e$, the expected one-step Lyapunov drift $\Delta^{\pi}(t),$ conditioned on the current virtual queue lengths $\tilde{Q}(t)$, under the operation of any admissible policy $\pi \in \Pi$ may be upper bounded as:
\begin{align}
    \Delta^{\pi}(t) & \equiv \mathbb{E}\Big(L(\tilde{Q}(t+1)) - L(\tilde{Q}(t))| \bm{\tilde{Q}(t)}\Big) \nonumber \\  
    &\leq B + 2\sum_{e \in E} \big(\tilde{X}_{e}(t)+ \tilde{Y}_{e}(t)\big)\mathbb{E}\big(A_e ^{\pi}(t) | \boldsymbol{\tilde{Q}(t)}\big) \nonumber \\
    &  \quad  \quad- 2\sum_{e \in E} \tilde{X}_{e}(t)\eta_e  - 2\sum_{e \in E} \tilde{Y}_{e}(t)\gamma_e,   \label{eqLyap}
\end{align}
where $B \equiv m(2A_{\max}^2+K_{\max}^2+\gamma_{\max}^2)$ is a finite constant. 
%Note that, in Eqn.\ \eqref{eqLyap}, we have used the fact that \[\mathbb{E}\big(K_e(t)|\bm{\tilde{Q}}(t)\big)= \mathbb{E}(K_e(t)) = \eta_e. \]
 \subsection*{A Drift Minimizing Routing Policy:} 
%We minimize the routing cost which is the positive term in the middle.
We now design a routing policy which minimizes the upper bound \eqref{eqLyap} on the one-step Lyapunov drift. By inspecting the terms on the bound, it is clear that the routing policy must choose the route for each packet to minimize the following routing cost:  
 $$ \text{RoutingCost}^{\pi} = \sum_{e \in E} \big(\tilde{X}_{e}(t) + \tilde{Y}_{e}(t)\big)A_e ^{\pi}(t).$$
 Using Eqn \eqref{yhnb}, we can express this cost as:
  $$ \text{RoutingCost}^{\pi} = \sum_{c \in \mathcal{C}}A^{(c)}(t)\sum_{e \in E} \big(\tilde{X}_{e}(t) + \tilde{Y}_{e}(t)\big)\mathds{1}(e \in T^{(c)}(t)),$$
 where $T^{(c)}(t) \in \mathcal{T}^{(c)}(t)$ and  $\mathcal{T}^{(c)}(t)$ is the set of all admissible routes for the packets belonging to the traffic class $c$. Decomposing the above cost function into distinct traffic classes, we see that the drift minimizing policy chooses routes for the packets in class $c$ at time $t$ by solving the following combinatorial optimization problem:
  \begin{align}
      T_{\text{opt}}^{(c)}(t) \in \underset{T^{(c)} \in \mathcal{T}^{(c)}(t)}{\arg\min}\sum_{e \in E} \big(\tilde{X}_{e}(t) + \tilde{Y}_{e}(t)\big)\mathds{1}(e \in T^{(c)}) \label{jnk}
  \end{align} 
  Let $W_e(t) \equiv \tilde{X}_e(t) + \tilde{Y}_e(t)$ be the sum of the  lengths of the virtual queues (consisting of both unencrypted and encrypted packets) for the edge $e$ at time $t$. Now consider an edge-weighted version of the graph $G$, where the weight of the edge $e$ is taken to be $W_e(t).$
 For different traffic types, the optimal route for each packet is chosen as follows:
 
  %on the graph $\mathcal{G}$, with edge weights as virtual queue lengths $\tilde{Q}$ of active edges:
 %\begin{tcolorbox}
  \begin{itemize}
  \item \textbf{Unicast:} The shortest $s^{(c)}- t^{(c)}$ path in the weighted-graph.
  \item \textbf{Broadcast:} The minimum-weight spanning tree (MST) with root $s^{(c)}$, in the weighted-graph.
  \item \textbf{Multicast:} The minimum-weight Steiner tree with root $s^{(c)}$ and covering all destinations $\mathcal{D}^{(c)}$ in the weighted-graph.  
  \item \textbf{Anycast:} The shortest of the $k$ shortest $s^{(c)}- t_i^{(c)}, 1\leq i \leq k$ paths in the weighted-graph.
\end{itemize}
%\end{tcolorbox}
For routing multicast traffic, we may use an efficient approximation algorithm for the Min-weight Steiner tree problem (such as the one described in \cite{byrka2010improved}), as solving the problem optimally is \textbf{NP-hard} for arbitrary graphs. For all other traffic classes, standard algorithms may be used for routing \cite{cormen2009introduction}. 

\subsection{Strong Stability of the Virtual Queues} \label{secrand}
We now show that the proposed TQD policy stabilizes the virtual queues in the network. 
\begin{theo} \label{main-thm}
 Under the TQD routing policy, the virtual queue process $\{\tilde{\bm{Q}}(t)\}_{t \geq 0}$ is strongly stable for any arrival rate vector $ \boldsymbol{\lambda} \in \textrm{int}(\overline { \boldsymbol{\Lambda}_{\omega}}),$ \emph{i.e.,}
%\begin{empheq}[box=\widefbox]{align}
\begin{eqnarray*}
    \limsup\limits_{T\rightarrow \infty} \frac{1}{T} \sum_{t=0}^{T-1}\sum_{e \in \text{E}}^{} \mathbb{E}\big(\tilde{X}_{e}(t)+\tilde{Y}_{e}(t)\big)< \infty.   
    \end{eqnarray*}
%\end{empheq}
\end{theo}
\begin{IEEEproof}
 Consider an arrival rate vector 
 $ \boldsymbol{\lambda} \in \textrm{int}(\overline{\boldsymbol{\Lambda}}_{\omega})$. From the definition of the set $\overline{\boldsymbol{\Lambda}}_{\omega}$ given by Eqns.\ \eqref{wde} and \eqref{rdde}, it follows that there exists a scalar $\epsilon > 0$ such that we can decompose the total arrival for each class $c \in \mathcal{C}$ into a finite number of routes, such that
\begin{align}
    \lambda_e =  \sum_{\substack{(i,c): e\in T_i^{(c)},\\ T_i^{(c)} \in \mathcal{T}^{(c)}}} \lambda_i^{(c)}\leq \omega_e -\epsilon, \quad \forall e \in E. \label{frt}
\end{align}
We now define an auxiliary stationary randomized routing policy $\pi_{\textbf{RAND}} \in \Pi$ such that the policy $\pi_{\textbf{RAND}}$ assigns an incoming packet from class $c$ the route $T_i^{(c)} \in \mathcal{T}^{(c)}(t)$ with probability $\frac{\lambda_i^{(c)}}{\lambda^{(c)}}, \forall i,c$. Hence, it follows that the expected number of packets that is routed along a path (or tree) that includes the edge $e$ is given by:
\begin{align} \label{fl-decomp}
    \mathbb{E}(A_e^{\pi_{\textbf{RAND}}}(t)) = \lambda_e =  \sum_{\substack{(i,c): e\in T_i^{(c)},\\ T_i^{(c)} \in \mathcal{T}^{(c)}}} \lambda_i^{(c)}, \quad \forall e \in E.
\end{align}
%
%As we are considering the wired topology here, we can say that in every slot all links are activated and thus under the policy each edge will have: 
%\begin{align}
%    \mathbb{E}(\gamma_e^{\pi_{\textbf{RAND}}}(t)) = \gamma_e
%\end{align}

Since the TQD policy minimizes the upper-bound to the drift expression in Eqn. \eqref{eqLyap} among the set of all feasible routing policies $\pi \in \Pi$, by  comparing it with the randomized policy $\pi_{\textbf{RAND}},$ we can write:
\begin{align}
        \Delta^{\pi_{\textbf{TQD}}}(t) &\leq B + 2\sum_{e \in E}\big(\tilde{X}_{e} (t) + \tilde{Y}_{e}(t) \big) \mathbb{E}\Big(A_e^{\pi_{\textbf{RAND}}}(t) | \boldsymbol{\tilde{Q}_e(t)}\Big) \nonumber \\
        &  \quad  \quad- 2\sum_{e \in E} \tilde{X}_{e}(t)\eta_e  - 2\sum_{e \in E} \tilde{Y}_{e}(t)\gamma_e.   \label{eqLyap2}
\end{align}
Using the fact that Randomized policy is memoryless, and hence, independent of the virtual queue lengths $\boldsymbol{\tilde{Q}_e(t)}$, substituting the expression \eqref{fl-decomp} into the above drift inequality simplifies to:
\begin{align*}
   % &= B + 2\sum_{e \in E}\big(\tilde{X}_{e} (t) + \tilde{Y}_{e} (t)\big) \mathbb{E}\big(A_e^{\pi_{\textbf{RAND}}}(t)\big) \\
  %  & \quad \quad - 2\sum_{e \in E}\tilde{X}_{e}(t)\eta_e - 2\sum_{e \in E}\tilde{Y}_{e}(t)\mathbb{E}\big(\gamma_e^{\pi_{\textbf{RAND}}}(t)\big) \\
   \Delta^{\pi_{\textbf{TQD}}}(t)   &\leq  B + 2\sum_{e \in E}\Big((\lambda_e - \eta_e)\tilde{X}_{e} (t) + (\lambda_e - \gamma_e) \tilde{Y}_{e} (t)\Big)\\
     &\leq  B + 2\sum_{e \in E}\Big(\big(\lambda_e - \min(\gamma_e,\eta_e)\big)\tilde{X}_{e} (t) \\
     &\quad \quad \quad \quad \quad  \quad \quad+ \big(\lambda_e - \min(\gamma_e,\eta_e)\big)\tilde{Y}_{e} (t) \Big)\\
    &= B + 2\sum_{e \in E}(\lambda_e - \omega_e)\big(\tilde{X}_{e} (t) + \tilde{Y}_{e} (t)\big)\\    
    &\leq B - 2\epsilon \sum_{e \in E}\big(\tilde{X}_{e} (t) + \tilde{Y}_{e} (t)\big),
\end{align*}
where we have used the inequality from Eqn. \eqref{frt}. Taking expectation of both sides w.r.t. the virtual queue lengths $\tilde{Q}(t)$, we can bound the expected drift at slot $t$ as:
\begin{align*}
    \mathbb{E}L(\boldsymbol{\tilde{Q}}(t+1)) - \mathbb{E} L(\boldsymbol{\tilde{Q}}(t)) \leq B - 2\epsilon\sum_{e \in E}\mathbb{E}\big(\tilde{X}_{e} (t) + \tilde{Y}_{e} (t)\big). 
\end{align*}
Upon summing the above inequality from $t = 0$ to $T-1$, dividing both sides by $T$ and upon realizing that $L(\bm{\tilde{Q}}(0))=0$ we have: 
\begin{align} \label{main-eq}
	\frac{\mathbb{E}L(\tilde{Q}(T))}{T} +  \frac{1}{T}\sum_{t=0}^{T-1}\sum_{e \in E}^{} \mathbb{E}(\tilde{X}_{e} (t) + \tilde{Y}_{e} (t)) \leq \frac{B}{2\epsilon}.
\end{align}
Finally, using the fact that $L(\boldsymbol{\tilde{Q}}(T)) \geq 0$, we get
\begin{align*}
    \frac{1}{T}\sum_{t=0}^{T-1}\sum_{e \in E}^{} \mathbb{E}\big(\tilde{X}_{e} (t) + \tilde{Y}_{e} (t)\big) \leq \frac{B}{2\epsilon}.
\end{align*}
Taking $\limsup$ on both sides we get that
\begin{align*}
    \limsup\limits_{T\rightarrow \infty} \frac{1}{T} \sum_{t=0}^{T-1}\sum_{e \in E}^{} \mathbb{E}\big(\tilde{X}_{e}(t)+\tilde{Y}_{e}(t)\big)< \infty,
\end{align*}
which shows that both of the virtual queue processes $\{\bm{\tilde{X}}(t)\}_{t \geq 1}$ and $\{\bm{\tilde{Y}}(t)\}_{t \geq 1}$ are strongly stable. 
\end{IEEEproof}
\paragraph*{Discussion} It is clear that, with the above TQD architecture, the proof of Theorem \ref{main-thm} goes through even when we do not store the keys from the past, \emph{i.e.} the freshly-generated keys at each slot are used for encrypting the packets for that slot only and the residual keys (if any) are discarded at the end of the slots ($k_e(t)=0, \forall t,e$). This is obviously a wasteful way of operating the system, but it does not affect the throughput-optimality of the TQD policy. Nevertheless, one advantage of this scheme is that we can now operate the system with \emph{zero-sized} key banks and discard stale (and potentially vulnerable) keys \emph{without} losing capacity. This observation is surprising in the context of the counterexample in Section \ref{secD}, where we showed that any policy $\pi^{\textrm{single queue}}$ with a  single-queue architecture with excess key discarding is provably \emph{not} throughput-optimal. 
%See section \ref{sims} for a numerical evaluation of its performance. 
%with the storage-based scheme. 

%\subsection{Rate Stability of the Virtual Queues}
\subsection{Stability of the Physical Queues} \label{phy_rate_stability}
The physical queues naturally obey precedence constraints and have a more complex dynamics than the virtual queues. In the following, we argue that the physical queues $\{\bm{X}(t)\}_{t \geq 1}$ and $\{\bm{Y}(t)\}_{t \geq 1}$ are also stable. 

\paragraph{Stability of the $\{\bm{X}(t)\}_{t \geq 1}$ process} Note that the number of keys generated at each slot for serving the virtual queue $\tilde{X}_e(t)$ and the physical queue $X_e(t)$ are identical for all edges $e\in E$ and time slot $t$. Since the excess keys are indefinitely stored in the key banks and, since the packet arrivals are counted in the virtual queue of unencrypted packets $\tilde{X}_e$ \emph{before} they actually arrive in the corresponding physical queue $X_e$, it readily follows that 
\begin{eqnarray} \label{unen}
X_e(t) \leq \tilde{X}_e(t), ~\forall e, t. 	~\textrm{a.s.}
\end{eqnarray}
Hence, from the virtual queue stability Theorem \ref{main-thm}, it follows that the physical queues $\bm{X}(t),$ consisting of the unencrypted packets, are strongly stable. 
\paragraph{Stability of the $\{\bm{Y}(t)\}_{t \geq 1}$ process}
Since, unlike the quantum keys, the cumulative unused services of the physical links cannot be stored for future use, it is not possible to derive a pairwise comparison inequality similar to \eqref{unen} for the downstream queues $\bm{Y}(t)$ and their virtual counterparts. %Our proof of the stability of the physical queues containing the encrypted packets (\emph{i.e.,} $\{\bm{Y}(t)\}_{t \geq 1}$) follows along the same line as given in \cite{sinha2017optimal}. 
Hence, we study the sample path behavior of the queueing processes to conclude their stability.  
Using the fact that the virtual queue processes $\{\bm{\tilde{X}}(t)\}_{t \geq 1}$ and  $\{\bm{\tilde{Y}}(t)\}_{t \geq 1}$ are non-negative, and $L(\bm{\tilde{Q}}(t)) \geq \tilde{Y}_e^2(t) , \forall e \in E,$ (viz. Eqn.\ \eqref{lyap}) from Eqn.\ \eqref{main-eq} we have $\mathbb{E}\tilde{Y}_e^2(T) \leq \frac{BT}{2\epsilon}, \forall T \geq 1. $ Furthermore, since the number of packet  arrivals at a slot and the link capacities are bounded, we see that the conditions of Lemma 3.2 of \cite{neely2012stability} are satisfied. %Hence, we have 
%\begin{align*}
%	\frac{\tilde{Y}_e(T)}{T} \leq \frac{A(T)}{T} \leq A_{\max}, 
%\end{align*}
%where we recall $A(T)$ denote the total number of packets injected to the network up to time $T$. Consider a subsequence $t_k$ such that Hence, using the Bounded Convergence Theorem, we have 
%\begin{eqnarray*}
%\mathbb{E}\bigg(\limsup_{T\to \infty} \frac{\tilde{Y}_e(T)}{T} \bigg)	= \limsup_{T \to \infty} 
%\end{eqnarray*}
%the r.v. $\nicefrac{\tilde{Y}_e(T)}{T}$ is dominated by the r.v. $\nicefrac{A(T)}{T},$ where $A(T)$ denotes the total number of packets that arrive to the network up to the time $T$. 
%using 
%
%the Dominated Convergence Theorem, it follows that 
%\begin{eqnarray*}
%\mathbb{E}\bigg(\limsup_{t \to \infty} \frac{\tilde{X}_e(t)+\tilde{Y}_e(t)}{t}\bigg)	= \limsup_{t \to \infty} \frac{\mathbb{E}\big( \tilde{X}_e(t)+ \tilde{Y}_e(t) \big)}{t}.
%\end{eqnarray*}
%
%Strong Stability Theorem (Theorem 2.8) from \cite{neely2010stochastic}, it follows that the Strong Stability of the virtual queues implies the rate stability of the virtual queues as well. 
Hence, under the TQD policy, we have for any $\lambda \in \text{int}(\boldsymbol{\overline{\Lambda}}_{\bm{\omega}})$: 
%\begin{empheq}[box=\widefbox]{align}
    %\lim_{t \to \infty} \frac{\tilde{X}_e(t)}{t} = 0, \quad \forall e \in E, \quad \quad \text{w.p.}~ 1 \\
    \begin{align} \label{rate-stability}
       \lim_{t \to \infty} \frac{\tilde{Y}_e(t)}{t} = 0, \quad \forall e \in E, \quad \quad \text{a.s.} 
    \end{align}
%\end{empheq}
Next, using an appropriate packet scheduling policy for the encrypted packets for the outgoing physical links (\emph{e.g.,} the Nearest to Origin policy \cite{gamarnik1999stability}), it can be shown that the rate stability condition of the virtual queues \eqref{rate-stability} implies the rate stability of the physical queues for the encrypted packets as well, \emph{i.e.,}
   \begin{align}
     \lim_{t \to \infty} \frac{\sum_{e \in E }{Y}_e(t)}{t} = 0,  \quad \quad \text{w.p.} ~1.
     \end{align}
%\end{empheq}
%We will use this condition to prove our throughput optimality.
We refer the readers to \cite{sinha2017optimal}, Theorem 3 for a detailed proof using adversarial queueing theory, which goes through without any modification. 
From the above, it immediately follows that the TQD policy is throughput-optimal. We give a formal proof of this result in Appendix \ref{tput}.

\section{Extension to Heterogeneous Networks with Multiple Security Levels}\label{multilevel}
\paragraph*{Setup} So far in this paper, we have considered an idealistic QKD network where \emph{all} transmitted packets need to be encrypted with the quantum keys and \emph{all} physical links are equipped with an overlay QKD module. However, in practice, depending on the required degree of confidentiality, multiple security levels may need to be supported. Furthermore, in a large heterogeneous network, such as the Internet, only a small fraction of the links possess overlay QKD modules. In this setting, consider a scenario where a group of users, denoted by $\mathcal{S}^*,$ intend to communicate confidential messages among them over a large heterogeneous network. In this case, only the packets originating from the users in the set $\mathcal{S}^*$ are required to be encrypted with the quantum keys, whereas the standard application layer encryption protocol suffices for the rest of the packets. We now describe an extension of the proposed TQD architecture, called Extended TQD (\emph{e}-TQD), that achieves the secure capacity region in this setting. 

\paragraph*{Extended-TQD (e-TQD)} Denote the set of all physical links possessing an associated overlay QKD module by $E_S \subseteq E.$ Clearly, packets from the highest security group $\mathcal{S}^*$, which need to be encrypted with the quantum keys before each transmission, can be routed only over the links in the set $E_S$. On the other hand, packets originating from sources other than the set $\mathcal{S}^*$ may be routed over any subset of links in the network. This observation suggests the following modifications to Algorithm \ref{tqd-algo} (see Algorithm \ref{tqd-algo2} and Algorithm \ref{tqd-algo3} in Appendix \ref{e-TQD-section} for the pseudocode).
  \begin{figure}
          \centering
          \begin{subfigure}[b]{1\textwidth}
       \includegraphics[height=0.1\textwidth]{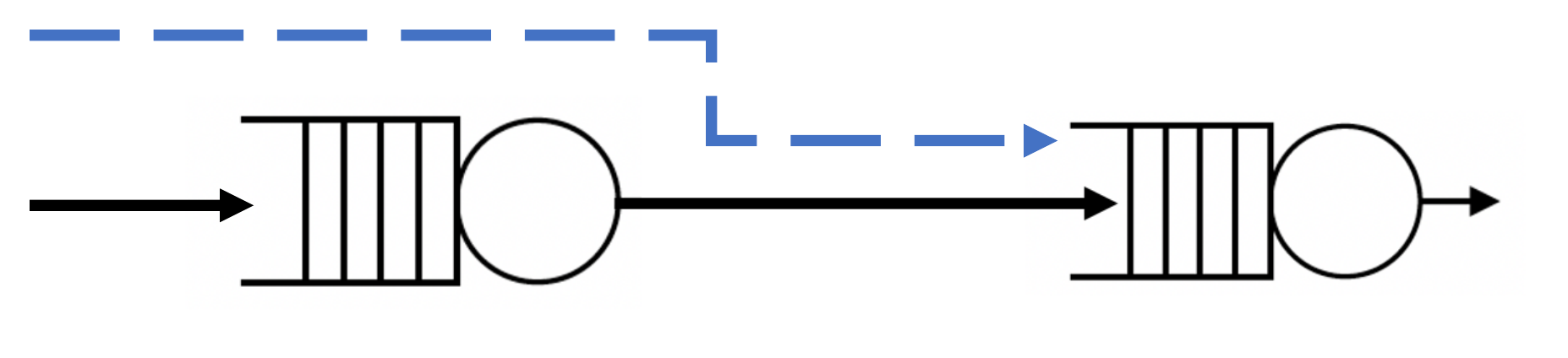}
    %   \put(-220,-10){\textrm{\small{\textcolor{black}{Unencrypted packets}}}}
     %  \put(-215, -20){\textrm{\small{(waiting for keys)}} } 
       \put(-185, -4){\textrm{\small{$X_e(t)$}} }
       \put(-65, -4){\textrm{\small{$Y_e(t)$}} }
       \put(-162, 3){\textrm{\scriptsize{(Encryption)}}}
       \put(-45, 3){\textrm{\scriptsize{(Transmission)}}}
       \put(-224, 12){\textrm{\footnotesize{Packets}}}
       \put(-224, 4){\textrm{\footnotesize{from $\mathcal{S}^*$}}}
       \put(-224, 52){\textrm{\footnotesize{Packets from other sources}}}
       \put(-110, -8){\footnotesize{(a)}}
       \end{subfigure}
      % \put(-90,-10){\textrm{\small{\textcolor{black}{Encrypted packets}}}}
       %\put(-135, -20){\textrm{\small{(waiting due to the limited link capacity)}}}
       \begin{subfigure}[b]{1\textwidth}
       \hspace{90pt}
       \includegraphics[height=0.08\textwidth]{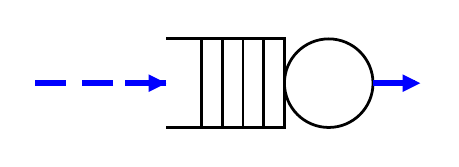}
       \put(-200, 22){\textrm{\footnotesize{Packets from the sources}}}
       \put(-200, 15){\textrm{\footnotesize{in $\mathcal{S}\setminus{\mathcal{S}^*}$ that do not require}}}
        \put(-200, 8){\textrm{\footnotesize{quantum encryption}}}
       \put(-70,-5){\footnotesize{(b)}}
       \end{subfigure}
          \caption{\footnotesize{(a) Depicting a link $e \in E_S,$ possessing an overlay QKD module, which serves both secured (from the sources $\mathcal{S}^*$) and unsecured (from the sources $\mathcal{S}\setminus \mathcal{S}^*$) packets. Packets which need not be encrypted with quantum keys skip the first queue and join the second queue directly. (b) Depicting a link $e \in E\setminus E_S$ without an overlay QKD module. This link only serves packets from the sources in the set $\mathcal{S} \setminus \mathcal{S}^*$ that do not require quantum encryption. } }
                    \label{fig:tqd2}
 \end{figure}
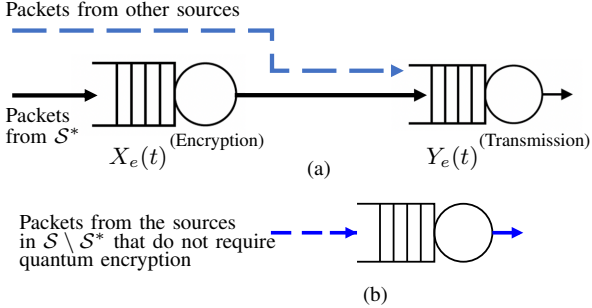
 The route of any packet, originating from some source in the set $\mathcal{S}^*$, is selected by computing the shortest path in the \emph{induced graph} $\mathcal{G}(V,E_S)$. Clearly, the induced subgraph contains only those edges possessing an overlay QKD module. On the other hand, packets originating from the sources in the set $\mathcal{S}\setminus \mathcal{S}^*,$ that do not require quantum encryption, skip the encryption queue $X_e$ for any edge $e \in E$. These packets directly join the downstream queue $Y$'s and wait for the transmission via the physical links. The links $e \in E\setminus E_S$ that do not possess overlay QKD modules, do not maintain the $X_e$ queues and maintain only the $Y_e$ queues. The virtual queue lengths are updated accordingly. See Figure \ref{fig:tqd2} for an illustration.
  %and Section \ref{e-TQD-section} in the Appendix for a complete description of the \emph{e}-TQD algorithm. 
  Using arguments similar to the proof in Theorem \ref{main-thm}, it can be shown that the proposed \emph{e}-TQD policy is throughput-optimal in this generalized multi-level security setting. We skip the proof due to space constraints. Numerical simulation results for the \emph{e}-TQD policy are given later in Section \ref{e-tqd-sim}.
 \begin{figure*}[ht]
\begin{subfigure}{.49\textwidth}
  \centering
  % include second image
%   \includegraphics[width=.8\linewidth]{image_file_name}  
   \includegraphics[width=0.9\textwidth]{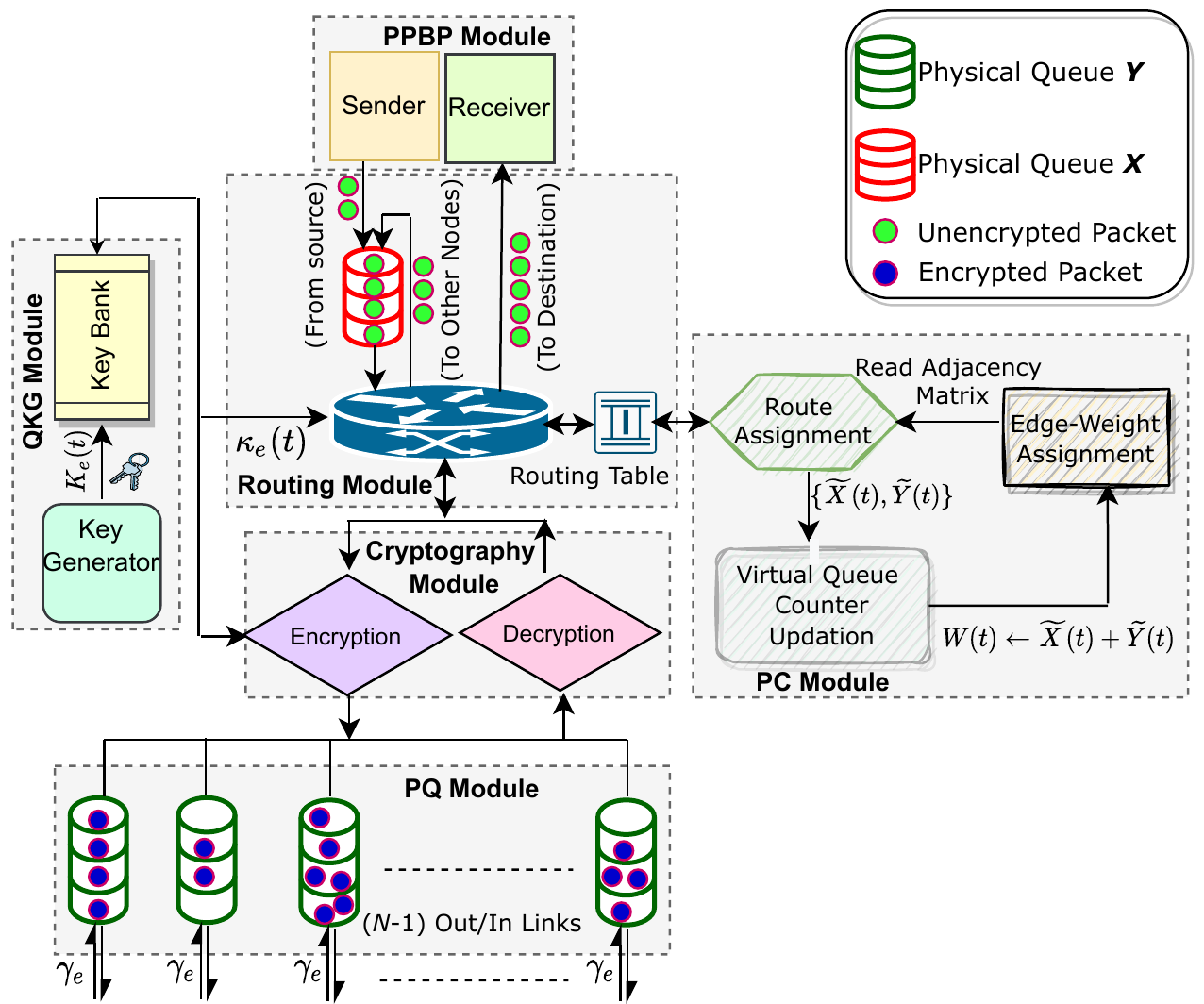}
  \caption{\footnotesize{Software architecture of the \textsc{TQDSimModule} node.}}
%   The node is made up of many physical modules that execute distinct network operations.
  \label{fig:sub-second}
\end{subfigure}
\begin{subfigure}{.49\textwidth}
  \centering
  % include first image
  \includegraphics[width=0.6\textwidth]{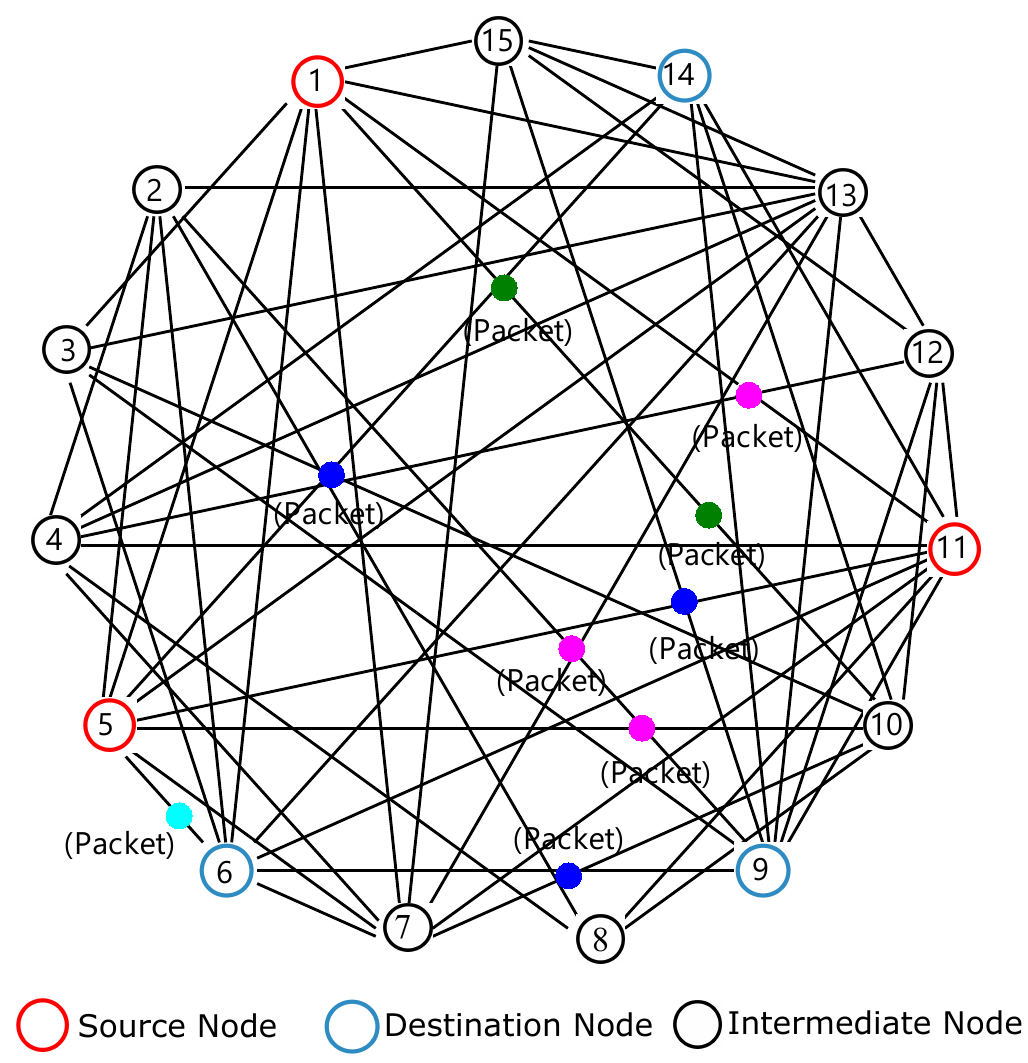}
  \caption{\footnotesize{A snapshot of the output from the \textsc{TQDSimModule} during simulation. For unicast traffic, the source and destination pairs are $(1, 9), (5, 14)$ and $(11, 6)$ respectively. For broadcast traffic, the source nodes are $1, 5,$ and $11$.}}
  \label{fig:sub-first}
\end{subfigure}
\caption{Depicting the \textsc{TQDSimModule}}
\label{fig:main_fig}
\end{figure*}

\section{Numerical Simulation} \label{sims}
\subsection{TQD Simulation Module (\textsc{TQDSimModule})} 

%As QKD technology development advances and gets more complicated, the need for a highly accurate and scalable simulation model rises. It is critical to assess the practical viability of every theoretical breakthrough in order to foresee challenges in its implementation. Furthermore, since QKD links necessitate optical and Internet connections between network nodes, a full test-bed with several network hosts and links would be prohibitively expensive to setup. Taking into account all the aforementioned points, in this section, we develop a TQD policy-based simulation module named \textsc{TQDSimModule} for simulating the TQD policy in the QKD networks.

%Furthermore, the QKD link requires optical and Internet connections between network nodes, setting up a full test-bed with several network hosts and links to check and verify the TQD routing protocol would be prohibitively expensive.

% \subsection{\textsc{TQDSimModule} Design}
The simulator used in this paper, which we name \textsc{TQDSimModule} \footnote{Source code available online at \cite{github}.}, is built on top of the state-of-the-art OMNeT++ network simulator platform \cite{virdis2019recent, varga2010omnet++}. OMNeT++ is a popular discrete event simulator written in C++. Previously, it has been successfully used in simulating queueing networks, wireless and ad-hoc networks, peer-to-peer networks, optical switches, and storage area networks. Table~\ref{tab:simulation_setup} describes the hardware and software configurations that we use in our numerical experiments. The values of the principal parameters used in the simulations are given in Table~\ref{tab:simulation_setup_2} and \ref{tab:PPBP_GEN_PARA}. A schematic of the software architecture of the \textsc{TQDSimModule} is shown in Figure \ref{fig:sub-second}. The simulator comprises of six major components, which we briefly describe below. 
%In the table, we have defined a parameter $\alpha, 0\leq \alpha \leq 1,$ that represents the degree of connectivity. 

%It can have a value ranging from 0 to 1, with 0 being a wholly disconnected network and 1 representing a fully linked network (i.e., a mesh network). For simplicity, we have considered a uniform propagation delay of 0.1ms between the adjacent nodes.

\begin{table}[h]
\caption{\small{System Configuration used for Simulation}}
\label{tab:simulation_setup}
\centering
%\resizebox{\linewidth}{!}{%
\begin{tabular}{ll}
\toprule
%\hline
Description & Details \\
\midrule
Operating System       & Ubuntu 20.04 LTS             \\ 
Processor              & Intel Core i5 7th Generation \\ 
Memory                 & 16~GB                         \\ 
Compiler               & gcc                          \\ 
Simulation Environment & OMNeT++ 5.6.2                \\
Simulation Script        & Cmdenv, Tcl/Tkenv            \\ 
\bottomrule
\end{tabular}
%}
\end{table}

%Figure~\ref{fig:main_fig} is the arbitrary network topology considered for simulating the TQD policy.  Figure~\ref{fig:sub-first} is the snapshot of the \textsc{TQDSimModule} during simulation, and Figure~\ref{fig:sub-second} shows the internal architecture of an individual node defined in the \textsc{TQDSimModule}. Each node is composed of multiple modules, each of which performs a specific function, as described below. 

\begin{table}[h]
\caption{\small{Simulation Parameters}}
\label{tab:simulation_setup_2}
%\resizebox{\linewidth}{!}{%
\centering
\begin{tabular}{ll}
\toprule
Parameters             & Value    \\                        %\\ \hline
\midrule
Number of Nodes ($N$)        & 150                               \\
Probability of connectivity ($p$) & 0.3                              \\
Duration of a time slot & 0.25 ms \\
Link Capacity ($\gamma_e$)          & 1 packet/time slot               \\
Maximum Queue Capacity ($Q_c$)   & 10, 000 packets                    \\
Mobility Model         & None                             \\
Propagation Delay      & 0.025~ms                                                       \\
Simulation Time        & $10^5$ time-slots \\
Simulation Style       & Cmdenv-express-mode              \\ 
\bottomrule
\end{tabular}%
%}
\end{table}

\subsubsection{Poisson Pareto Burst Process (PPBP) Module}
 Analysis of a series of network-layer traces has established that real network traffic exhibits self-similarity, \emph{i.e.,} its statistical behaviour remains invariant across multiple time scales \cite{leland1994self, crovella1997self}. 
 %Because data or video traffic is scale-invariant, most performance studies employing traditional Markovian traffic models do not represent the fractal nature of network traffic. 
 Since traffic models have profound implications for the performance of routing policies, it is instructive to test the proposed algorithms with realistic packet arrival models. 
  % In order to achieve an accurate and realistic performance study, it is important to mimic the system's behaviour with appropriate traffic injected into the network. In the context of telecommunications, the first traffic model, based on the Poisson arrival process, was established, where call arrivals could be regarded independent and identically distributed. However, the Poisson traffic model was shown to be unsuitable for simulating bursty traffic.
Poisson Pareto Burst Process (PPBP) is a widely used traffic model that emulates the statistical behavior of real-world network traffic~\cite{ammar2011new}. In our experiments, we use PPBP as a traffic source generator for injecting bursty traffic into the network. The parameters we use to produce the PPBP traffic are listed in Table~\ref{tab:PPBP_GEN_PARA}.

% This module has two functional blocks, viz., key generator and key storage bank. Key generator randomly generates private symmetric quantum keys between the neighbouring nodes at every slot. All the unused keys in the current slot is stored in the key bank to be used for next slot. The key bank shares the information of available keys with the routing module and cryptography module for both encryption of new packets and decryption of incoming packet from other nodes.
\begin{table}[h]
\caption{Traffic generation parameters}
\label{tab:PPBP_GEN_PARA}
\centering
%\resizebox{\linewidth}{!}{%
\begin{tabular}{ll}
%\hline
\toprule
Parameters                          & Details                             \\ 
\midrule
Traffic generation model            & PPBP \\
Traffic type                        & Bursty                              \\
Maximum number packets/burst        & 5000                                \\
Minimum number packets/burst        & 1                                \\
Sleep time       & 25 time-slot                                \\
Burst time       & 5 time-slot                                \\
Hurst parameter                     & 0.8                                 \\
Pareto shape parameter (ON-Period)  & 1.4                                 \\
Pareto shape parameter (OFF-Period) & 1.2                                 \\
Packet size                         & 512~B                  \\ 
\bottomrule
\end{tabular}
%}
\end{table}

%  \begin{figure}[h]
% \centering
% \includegraphics[width=0.45\textwidth,height=0.28\textheight]{TQD_Diagram.drawio}
% \caption{The internal architecture of a single TQDSimModule node. The node is made up of many physical modules that execute distinct network operations.}	\label{TQDSimModule_Ref}
% \end{figure}

\subsubsection{Quantum Key Generation Module (QKG)}
Two major functions of the QKG module are the generation of \emph{new} keys at each time slot and the storage of the residual keys in the key banks. At each slot $t$, a random number $K_e(t)$ of private symmetric quantum keys are generated over the QKD link $e$, where $K_e(t) \sim \textrm{Poisson}(\eta_e)$. 
%OMNeT++ supports a wide range of random number generators (RNGs) for both continuous and discrete distributions. 
%Since, key generation rate follows Poisson process, we have used an in-built function \emph{poisson(lambda)} available in the simulation library. 
The key-generation rate is fixed for each edge during the network initialization such that the mean rate $\eta_e$ is uniformly distributed in the range $[0.2-1]$. 
%It is worth noting that once a specific lambda value is set for an edge, it won't change until the simulation time is not reached. 
At the start of each slot, the routing module shares information about the total number of available keys $\kappa_e(t)$ with the cryptography module. The residual keys from the current slot are stored in the key bank for future use. 

\subsubsection{Policy Controller Module (PC)} This module is responsible for maintaining the weights in the graph $\mathcal{G}(V,E)$. In particular, it is responsible for updating the virtual queue counters ($\tilde{\bm{X}}(t)$ and $\tilde{\bm{Y}}(t)$), assigning edge-weights ($\boldsymbol{W}(t)$), and computing the minimum-weight routes for each incoming packets. The routing module communicates with the PC module for updating the routing table.

% The TQD module selects the optimal route for each incoming packet received from the PPBP module. The selection of optimal route is based on the network graph weight and traffic class of the packet.
\subsubsection{Cryptography Module}
All cryptographic tasks, such as symmetric key encryption, decryption, and authentication are performed by this module. This module interfaces with the routing module and the QKG module. Packets are encrypted if sufficiently many keys are available on the key banks. Otherwise, packets are queued in the physical queue $X_e$ until the quantum keys become available.

\subsubsection{Physical Queue (PQ) Module}
This module has multiple physical queues $Y_e$ for enqueuing the encrypted packets either received from the routing module of the same node or from the adjacent nodes. 
%The number of physical queues is equal to the number of outgoing edges from the node, which means one queue for each edge. 
In order to analyze the total packet drops in the network, the queue's capacity is assumed to be finite in the simulations. 

\subsubsection{Routing Module}
All communication among different modules takes place through the routing module. The physical queue $X_e$, which stores unencrypted packets, is a part of this module. The encrypted packets received from the upstream nodes are first decrypted by the cryptography module and then either delivered to the sink (if this is the destination node) or sent to the physical queue $X_e$ for the next hop encryption. 
\subsection{Simulation Results}
We now numerically compare the performance of the TQD policy with other standard benchmarks for different types of traffic. The policies are simulated on an Erdos-Renyi random topology having $N=150$ nodes such that any two nodes are connected independently with probability $p=0.3$.
% with 15 unicast flows.
%\section{\textcolor{red}{Simulation Results}} \label{secH}
\subsubsection{TQD with Unicast traffic}
\begin{figure}[h]
\centering
\includegraphics[width=2.7in,height=1.9in]{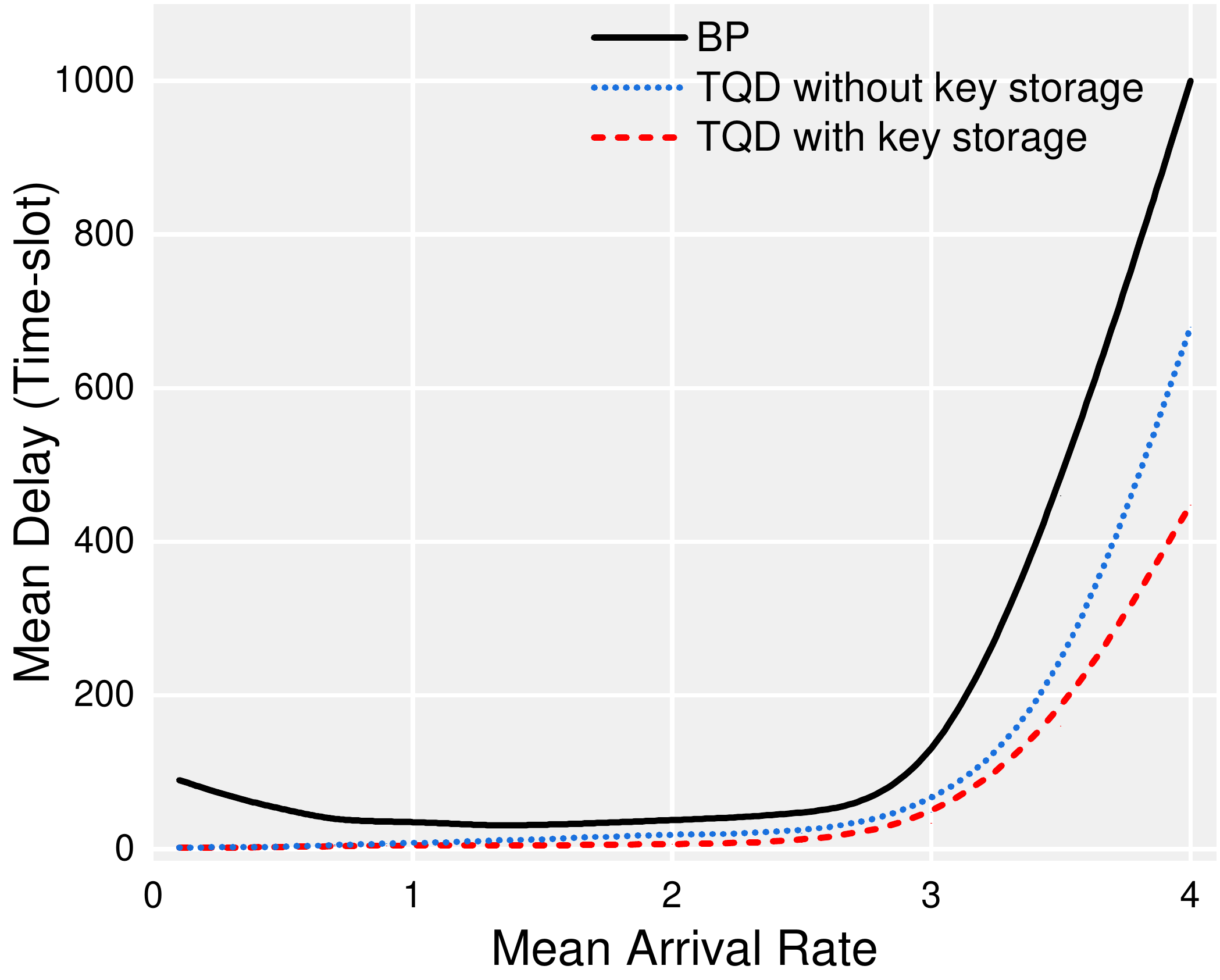}
\put(-75,130){\scriptsize{\cite{zhou2019security}}}
\caption{\small{Mean packet delay comparison between the TQD policy (with and without key storage) and the Backpressure policy \cite{zhou2019security}.}}
\label{Result_1}
\end{figure}

 The source-destination pairs for $15$ unicast flows are selected uniformly at random from the set of all $N=150$ nodes. The relevant parameter settings used in our simulation are given in  Table~\ref{tab:simulation_setup},~\ref{tab:simulation_setup_2} and~\ref{tab:PPBP_GEN_PARA}. In the simulation, we compare the performance of the proposed \textit{Tandem Queue Decomposition} policy (with and without the key-storage) with the Backpressure-based QKD routing policy proposed recently in \cite{zhou2019security}.
Figure \ref{Result_1} shows the variation of the mean packet delay as a function of the arrival rate $\lambda$. Hence, it follows that the TQD policy (with or without key storage) clearly outperforms  the Backpressure policy in terms of the mean packet delay. As argued before, the TQD policy without key storage is throughput-optimal, but due to the discarding of residual quantum keys, it performs poorly compared to its key storage variant. From the plot, it can also be observed that the relative performance gain of the TQD policy compared to the Backpressure policy is more pronounced, especially at the lower and higher rate regimes. This is because, at lower arrival rates, the congestion gradients, which form the basis of the BP policy, are small. As a result, the average number of hops a packet traverses through the network before reaching its destination becomes large, which leads to excessive delays. On the other hand, for higher arrival rates, the TQD policy is more efficient than the BP policy, which, by design, stabilizes the number of residual keys.

%\begin{figure}[h]
%\centering
%\includegraphics[width=0.3\textwidth]{UnicastSim.png}
%\caption{Delay performance comparison of \textbf{TQD} (with and without key storage) and \textbf{Back pressure} policy in the unicast setting}
%\label{unic}
%\end{figure}

%\begin{figure}[h]
%\centering
%\includegraphics[width=2.7in,height=1.9in]{figures/HopCount_New_4.pdf}
%\caption{Mean hop count comparison between the \textbf{TQD} policy (with and without key storage) and the \textbf{Back pressure} policy.}
%\label{Result_2}
%\end{figure}

\begin{figure}[h]
\centering
\includegraphics[width=2.7in,height=1.9in]{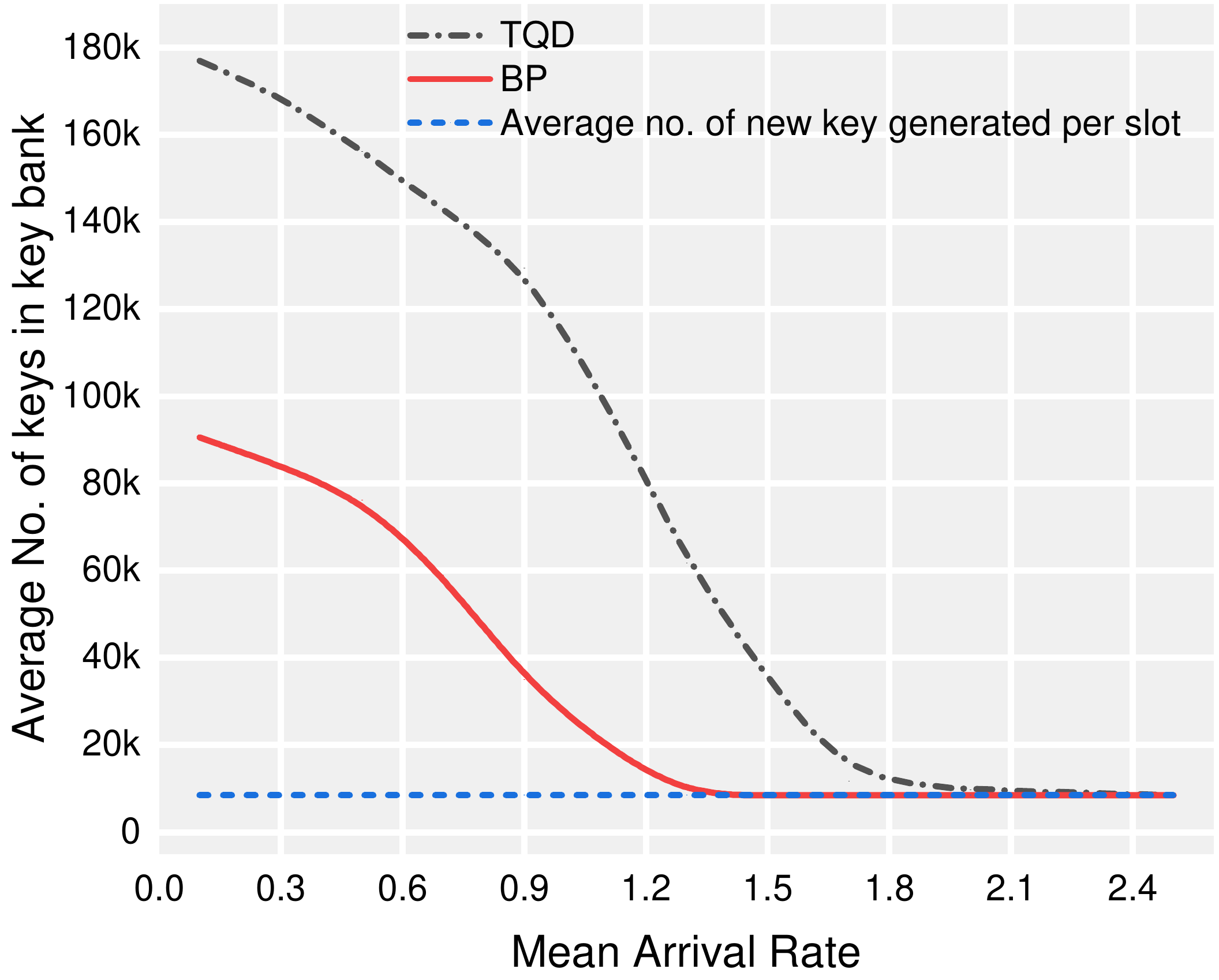}
\put(-108, 124){\scriptsize{\cite{zhou2019security}}}
\caption{\small{Average number of in-network residual keys available in the key banks for different arrival rates.}}
\label{Result_6}
\end{figure}

Figure~\ref{Result_6} compares the TQD and Backpressure policies in terms of the average number of in-network residual keys for different arrival rates. The plot shows that the TQD policy results in more in-network residual keys than the BP Policy. An abundance of the residual keys helps to mitigate the key availability constraints in the QKD networks and improve the end-to-end latency.  
%
%The contribution of packet delay owing to physical queues X and Y, as well as propagation delay, to the total end-to-end delay is shown in Figure~\ref{Result_5}. As expected, the key availability limitation is the biggest bottleneck of faster communication in QKD networks which is also evident from the plot~\ref{Result_5}, where physical queue X contributes the most to the total end-to-end delay. Physical queue Y offers the second most delay to the overall end-to-end delay, followed by propagation delay. Since our routing policy provides loop-free routing, the mean hop count of packet between source and destination is very low, resulting in very short propagation delay.
%\begin{figure}[h]
%\centering
%\includegraphics[width=2.7in,height=1.9in]{figures/Delay_Breakdown.pdf}
%\caption{\small{Contribution of Physical queue $X$ and $Y$ delays and propagation delay to total end-to-end delay.}}
%\label{Result_5}
%\end{figure}

\subsubsection{TQD with Broadcast traffic}
\begin{figure}[h]
\centering
\includegraphics[width=2.7in,height=1.9in]{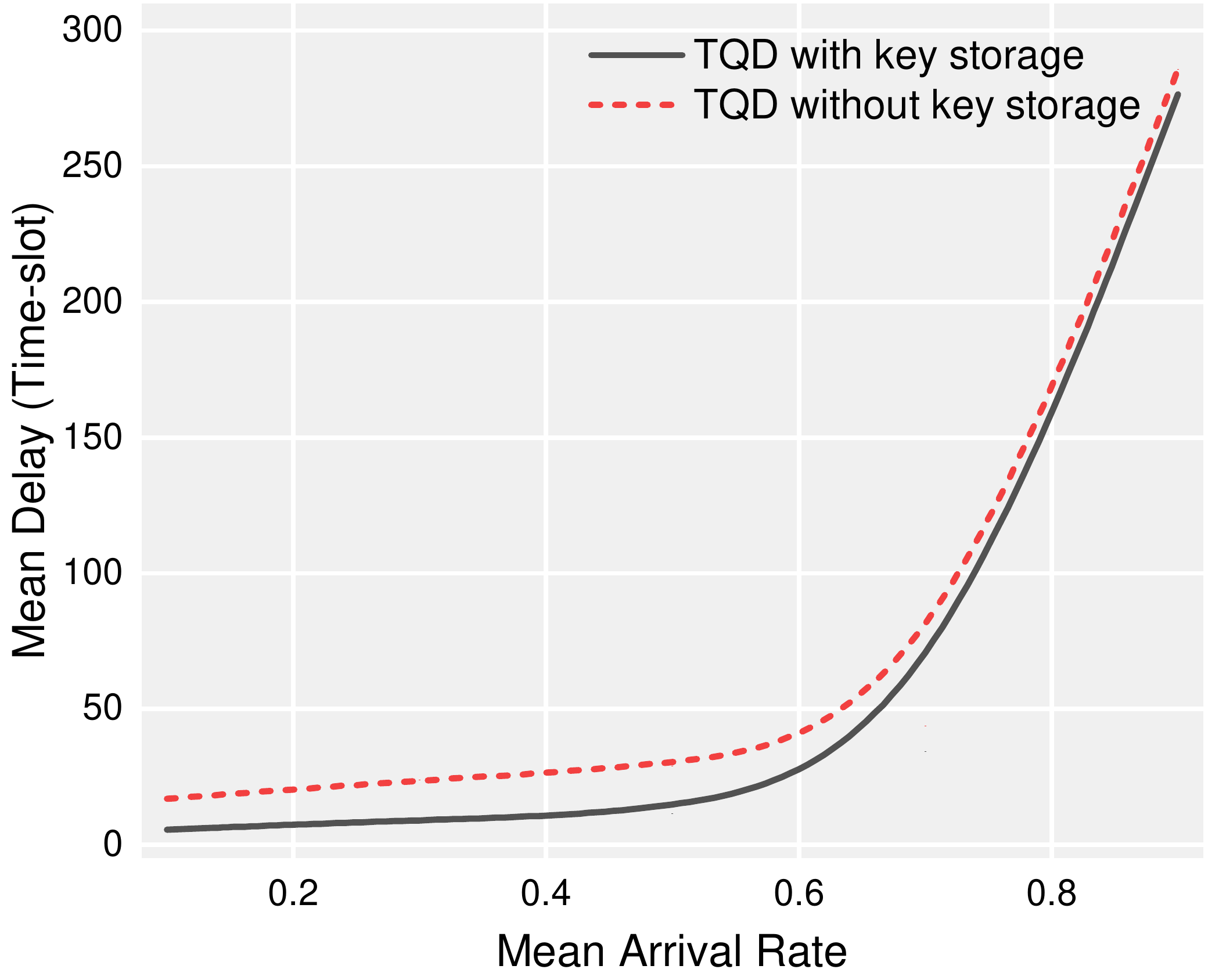}
\caption{\small{Delay Performance of the \textbf{TQD} policy (with and without key storage) for broadcast traffic}}
\label{Result_3}
\end{figure}

%For broadcast traffic, we use the same network topology and simulation settings as we use for unicast traffic.  
Figure \ref{Result_3} shows the difference in the average delay to broadcast packets between two variants of the TQD policy. We see that both policies are capacity-achieving, yet, unsurprisingly, the TQD policy with the key-storage variant outperforms its no-key-storage counterpart. However, the gap between the two variants reduces gradually on increasing the mean arrival rate. Since the algorithm proposed in the paper \cite{zhou2019security} cannot handle broadcast flows, its performance has not been shown on in this plot. 
%This is because with a very high load factor, all of the keys generated in a given slot will be used immediately. As a result, there will be no storage of unused keys in the key bank, leading to similar delay performance.

% \begin{figure}[h]
% \centering
% \includegraphics[width=0.37\textwidth]{Broadcast_plot.pdf}
% \caption{Delay Performance of the \textbf{TQD} policy for broadcast traffic}
% \label{twed}
% \end{figure}
\subsection{Performance of the e-TQD policy with multiple security levels} \label{e-tqd-sim}
In our final experiment, we consider network flows from sessions belonging to two distinct security levels - (A) sessions that require quantum encryption and (B) sessions that do not require quantum encryption. 
%Secured packets generated by highest security group $\mathcal{S}^*$, need to be encrypted with quantum keys before each transmission, and is routed over the links $E_S$ only. On the other hand, unsecured packets originated from sources other than $\mathcal{S}^*$, need not to be encrypted with quantum keys, and is routed using any subset of links from $E$. 
Packets belonging to group (A) are further categorized into two different sub-groups based on their relative priorities. 
%One group of secured packet has the highest priority, while the priority of the other group is equal to that of an unsecured packet.
We assume that the packets belonging to groups (A) and (B) are generated with equal probabilities. 

\begin{figure}[h]
\centering
\includegraphics[width=2.7in,height=1.9in]{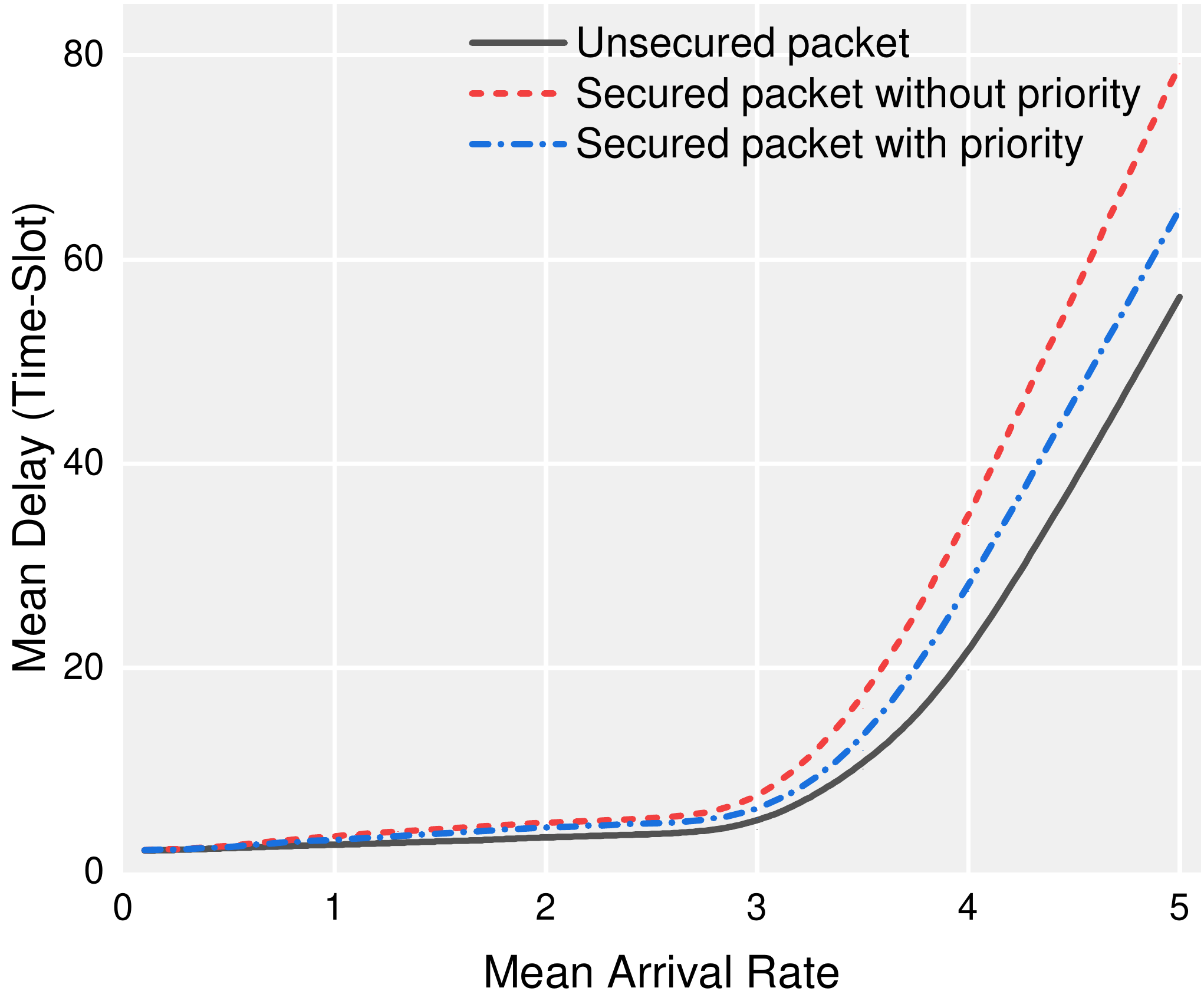}
\put(-45, 130){\scriptsize{\textrm{(Group (B))}}}
\put(-4, 120){\}}
\put(0, 121){\scriptsize{\textrm{(Group (A))}}}
\caption{\small{Delay performance of unencrypted packet and encrypted packets (with and without priority) under the e-TQD policy.}}
\label{Result_4}
\end{figure}
Figure~\ref{Result_4} shows the mean delay of the packets belonging to the groups (A) and  (B). It can be seen that packets belonging to the group (A) with the lowest priority level have the highest delay, followed by packets in the group (A) having the highest priority level, followed by packets in the group (B), which have the smallest average delay. This delay performance can be understood from the fact that, unlike the group (A) packets, the group (B) packets do not require quantum keys for their transmission. 

\section{Conclusion} \label{secI}
In this paper, we proposed a secure and provably throughput-optimal routing, scheduling, and key management policy for QKD networks carrying different types of traffic, including unicast, broadcast, multicast, and anycast. The policy is based on a simple Tandem Queue Decomposition architecture which effectively reduces the problem to a generalized network flow problem without the key availability constraints. We have investigated the proposed policies both analytically and with comprehensive numerical simulations. In the future, we plan to extend the proposed policies beyond the trusted node setting considered in this paper.  
% a policy Tandem Queue Decoupling which provides a throughput optimal solution for handling a mix of Unicast,
%Broadcast, Multicast and Anycast traffic in arbitrary wired networks and is empirically shown to have superior performance compared to the existing policies. The next step in the work would be to extend the policy to generalised wireless setting as well. 
\section{Acknowledgement} \label{ack}
This work is partially supported by the grant IND-417880 from
Qualcomm, USA and a research grant from the Govt.\ of India under the IoE initiative. 
%The computational results
%reported in this work were performed on the AQUA Cluster at
%the High Performance Computing Environment of IIT Madras. 

%\clearpage
\balance
\bibliographystyle{unsrt}  
\bibliography{Bibli.bib}

\clearpage
%\appendices
\nobalance
\appendix
\subsection{A brief description of the BB84 protocol} \label{bb84}
For completeness, we now briefly review the BB84 protocol originally invented by Charles Bennett and Gilles Brassard in $1984$ \cite{bennett2020quantum, shor2000simple}. 
The BB84 protocol defines a way of sharing secret keys over a quantum link between two nodes in which it is impossible to eavesdrop without disturbing the original transmission. This makes eavesdropping detectable with high probability by the communicating parties (traditionally denoted by Alice and Bob). 
The key idea is to encode each bit of the secret key into the polarization angle of a single photon. The polarization angles representing binary zero and one are together called a basis. The examples of two such bases are $0$, $90$ degrees (rectilinear (R))  and $45$, $135$ degrees (diagonal (D)). Because each vector of one basis has projections of equal length on all vectors of the other, these two bases are called conjugates. 
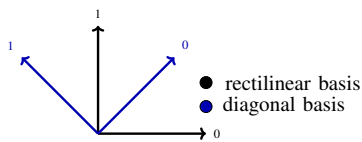
\begin{figure}[h]
	\centering
	\resizebox{5cm}{!}{	\colorlet{dRed}{red!80!black}
	\colorlet{dBlue}{blue!70!black}
	\colorlet{dGreen}{green!70!black}
	\colorlet{dPink}{purple!70!white}

	\def\ndr{3cm}
	\def\xfctr{3}
	\def\onovrrttwo{0.7071}
	
	\begin{tikzpicture}
	\draw[line width=2mm,->] (0,0) -- (0,\xfctr*\ndr) node[anchor=south,scale=3] {1};
	\draw[line width= 2mm,->] (0,0) -- (\xfctr*\ndr,0)node[anchor=west,scale=3] {0};
	\draw[line width= 2mm,->,color=dBlue] (0,0) -- (\xfctr*\ndr*\onovrrttwo,\xfctr*\ndr*\onovrrttwo)node[anchor=south west,scale=3] {0};
	\draw[line width= 2mm,->,color=dBlue] (0,0) -- (-\xfctr*\ndr*\onovrrttwo,\xfctr*\ndr*\onovrrttwo)node[anchor=south east,scale=3] {1};
	
	\draw[fill=dBlue](\xfctr*\ndr,0.75\xfctr*\ndr) circle (0.5cm) node[scale=5, xshift=1.3cm] {diagonal basis};
	%			\node[scale=3] at (\xfctr*\ndr*1.35,0.75\xfctr*\ndr) {diagonal basis};
	%			
	\draw[fill=black,yshift=2cm](\xfctr*\ndr,0.75\xfctr*\ndr) circle (0.5cm) node[scale=5, xshift=1.45cm] {rectilinear basis};
\end{tikzpicture}}
	\caption{Conjugate Bases}\label{fig:myfigure}
\end{figure}
Whenever a $\theta_1$ polarized light passes through a polarizer of a certain angle (say $\theta_2$), individual photons get either transmitted or absorbed with probability $p$ or $1-p$ respectively, where $p = \cos^2(\theta_1 - \theta_2)$. Thus,  complete deterministic information can only be available when the axis of the polarizer matches with that of the transmitting basis. In all other cases, the information will be lost.

\begin{figure}
 	\centering
 	\includegraphics[width = 0.5\textwidth]{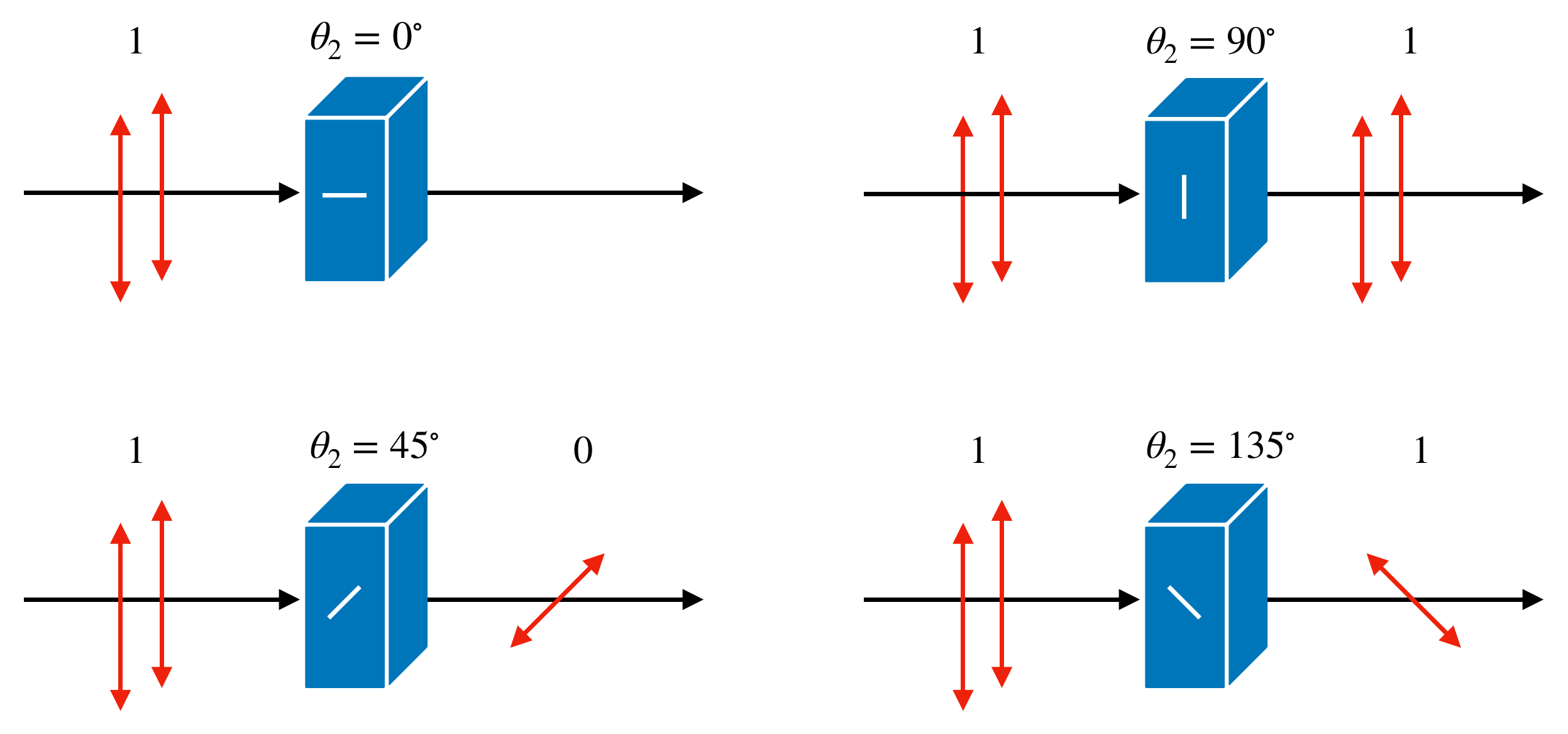}
 	\caption{Two vertically polarized photons with $\theta_1 = 90^{\circ}$ are passed through polarizers oriented at different angles. Deterministic outcomes are seen in top two cases ($\theta_2 = 0^\circ,90^{\circ}$), probabilistic absorption and transmission of photons is seen in the bottom two cases ($\theta_2 = 45^\circ,135^{\circ}$)}.
 \end{figure}  
 Alice takes a random bit string and transmits a train of photons whose polarization bases are chosen uniformly at random from the rectilinear and diagonal bases. Bob also chooses a sequence of bases independently and uniformly at random for measuring the polarization of each received photon. Any photon whose measurement polarization basis differs from the transmitted basis will produce binary zero or one with equal probabilities. In such cases, the original information stored in its polarization will be lost. This holds even for the eavesdropper. Therefore, Bob is required to confirm the correctness of his basis for each bit transmitted. Alice announces her choice of the bases over a classical authenticated channel. Bob discards all bits for which his choice of the basis was incorrect. Whenever a photon of one polarization state is measured with the same basis, it renders the same bit without any change. 
 Even if Alice confirms the bases, Bob cannot fully guarantee the correctness of the bits since any eavesdropping would have modified the transmitted polarisation angle, leading Bob to misinterpret a binary zero as one or vice versa. Therefore, Bob reveals a random subset of the key bits he received with the correct basis and confirms it with Alice. When Bob and Alice agree on the shared bits, it means that the channel is free from significant eavesdropping. If the shared bits do not match, the eavesdropper must have made a measurement that changed the quantum state. The whole key is discarded, and the procedure is repeated until Alice and Bob agree upon the revealed bits. These discussions over the classical channel do not compromise the secrecy of the remaining bits, as they are never revealed. The number of bits secretly shared at any time is random because of these polarization mismatches, noise, and  possible eavesdropping activity, which we model as a stochastic counting process with an appropriate rate.
\begin{table}[!h]
	\caption{Sample communication transcript between Alice and Bob using the BB84 protocol}
	\centering
	\renewcommand{\arraystretch}{1.2}%
	\newlength{\tcellwidth}
	\setlength{\tcellwidth}{2.5cm}
	\newcommand{\topcell}[1]{\begin{tabular}{@{}p{\tcellwidth} @{}} #1
	\end{tabular}}
	\begin{tabular}{@{} p{\tcellwidth}cccccccccc @{}} 
		\toprule 
		\multicolumn{10}{c}{\scshape Quantum Transmissions}  \\
		\midrule
		\topcell{Alice's random bits}  & 0 & 1 & 1 & 0 & 1 & 1 & 0 & 0 & 1 \\
		\midrule
		\topcell{Alice's random bases}  & D & R & D & R & R & R & R & R & D \\
		\midrule
		\topcell{Alice's photon polarization} & \rotatebox{45}{$\leftrightarrow$} & \rotatebox{90}{$\leftrightarrow$} & \rotatebox{135}{$\leftrightarrow$} & \rotatebox{0}{$\leftrightarrow$} & \rotatebox{90}{$\leftrightarrow$} & \rotatebox{90}{$\leftrightarrow$} & \rotatebox{0}{$\leftrightarrow$} & \rotatebox{0}{$\leftrightarrow$} & \rotatebox{135}{$\leftrightarrow$} \\
		\midrule
		\topcell{Bob's random bases} & R & D & D & R & R & D & D & R & D \\
		\midrule
		\topcell{Bob's Reception} &  1 & - & 1 & - & 1 & 0 & 0 & 0 & - \\
		\multicolumn{10}{l}{('-' indicates photon absorption due to incorrect angle of measurement)} \\
		\midrule
		\multicolumn{10}{c}{\scshape Classical Transmissions} \\
		\midrule
		\topcell{Alice tells her bases} &  R &  & D &  & R & D & D & R &  \\
		\midrule
		\topcell{Bob's confirmation} &  &  & \checkmark &  & \checkmark &  &  & \checkmark &  \\
		\midrule
		\multicolumn{10}{c}{Without eavesdropping}\\ \midrule
		\topcell{Shared Key} &   &  & 1 &  & 1 &  &  & 0 &  \\
		\midrule
		\topcell{Bob reveals some bits} &   &  &  &  & 1 &  &  &  &  \\
		\midrule
		\topcell{Alice's confirmation} &   &  &  &  & \checkmark &  &  &  &  \\
		\midrule
		\topcell{Remaining bits of secret key shared}& &  & 1 &  &  &  &  & 0 &  \\
		\midrule \multicolumn{10}{c}{With eavesdropping}\\ \midrule
		\topcell{Shared Key} &   &  & 1 &  & 1 &  &  & \textcolor{red}{1} &  \\
		\midrule
		\topcell{Bob reveals some bits} &   &  &  &  & 1 &  &  & 0 &  \\
		\midrule
		\topcell{Alice's confirmation} &   &  &  &  & \checkmark &  &  & \color{red}{$\times$} &  \\
		\multicolumn{10}{c}{DISCARD \& RETRANSMIT} \\
		\bottomrule
	\end{tabular} 
\end{table}
\color{black}
\subsection{Proof of the converse part of Theorem \ref{cap_ch}} \label{app1}
Consider any admissible arrival rate vector $\boldsymbol{\lambda} \in \boldsymbol{\Lambda}(\mathcal{G},\bm{\eta}, \bm{\gamma})$. By definition, there exists an admissible policy $\pi \in \Pi$ which supports the arrival vector $\boldsymbol{\lambda}$. Without any loss of generality, we may assume the policy $\pi$ to be stationary and the associated DTMC to be ergodic. Let $A^{(c)}_i(0,t)$ denote the number of packets belonging to class $c$ that have arrived at all of their destination(s) along the route $T_{i}^{(c)} \in \mathcal{T}^{(c)}$ up to time $t$. Recall that each packet is routed along one admissible route only. Thus we can say:
\begin{align} 
\sum_{T_{i}^{(c)} \in \mathcal{T}^{(c)}}A_i^{(c)}(0,t) = R^{(c)}(t), \label{eq0}
\end{align}
where $R^{(c)}(t)$ represents the number of distinct class-$c$ packets received by all destination nodes $\mathcal{D}^{(c)}$ under the action of the policy $\pi$, up to time $t$.
We also know that if $A^{(c)}(0,t)$ represents the total number of class-$c$ packet arrivals to the source $s^{(c)}$ up to time $t$, then:
\begin{align}
 A^{(c)}(0,t) \geq  \sum_{T_{i}^{(c)} \in \mathcal{T}^{(c)}}A_i^{(c)}(0,t),
 \label{eq:1}
\end{align}
as any packet that has finished its journey along some route $T_{i}^{(c)} \in \mathcal{T}^{(c)}$ by the time $t$, must have arrived at the source before that.
By dividing the inequality \eqref{eq:1} by $t$ and taking limit $t \to \infty$, we have:
\begin{align*}
    \lim_{t \to \infty} \frac{A^{(c)}(0,t)}{t} & \geq \liminf_{t \to \infty} \frac{1}{t} \sum_{T_{i}^{(c)} \in \mathcal{T}^{(c)}}A_i^{(c)}(0,t)\\
    &\stackrel{(a)}{=} \liminf_{t \to \infty}\frac{R^{(c)}(t)}{t}\\
    &\stackrel{(b)}{=} \lambda^{(c)}.
\end{align*}
The equality (a) holds from Eqn. \eqref{eq0} and the equality (b) holds from the definition \eqref{uiy} and the fact that our policy $\pi \in \Pi$ is claimed to securely support the arrival rate $\bm{\lambda}$.
By using SLLN we can say that 
$$\lambda^{(c)} = \lim_{t \to \infty} \frac{A^{c}(0,t)}{t}.$$
From this we can conclude that w.p. 1
\begin{align}
    \lim_{t \to \infty} \frac{1}{t} \sum_{T_{i}^{(c)} \in \mathcal{T}^{(c)}}A_i^{(c)}(0,t) = \lambda^{(c)}, \quad \forall c \in \mathcal{C} \label{eq:2}  
\end{align}
Using the fact that the policy $\pi$ is stationary and the associated DTMC is ergodic, we conclude that the time-average limits exist and they are constant a.s. For all $T_{i}^{(c)} \in \mathcal{T}^{(c)}$ and $c \in \mathcal{C}$, define
\begin{align}
    \lambda_i^{(c)} &\stackrel{\text{(def)}}{=} \lim_{t \to \infty} \frac{A_i^{(c)}(0,t)}{t}. \label{eq:f}  
\end{align}
Using Eqns. \eqref{eq:2} and \eqref{eq:f} we get
\begin{align} \label{eqp}
    \lambda^{(c)} = \sum_{T_{i}^{(c)} \in \mathcal{T}^{(c)}} \lambda_i^{(c)}.
\end{align}
The previous equation \eqref{eqp} proves Eqn. \eqref{wde} that there exists a non-negative flow decomposition of the incoming packets amongst the admissible routes.

For the second part of proof, we consider an edge $e \in E$ in the graph $\mathcal{G}$. Let $A_e(0,t)$ be the number of packets that have crossed edge $e$ till time $t$ under the action of the policy $\pi$. We have that:
\begin{align}
    \sum_{\mathclap{\substack{(i,c): e\in T_i^{(c)},\\ T_i^{(c)} \in \mathcal{T}^{(c)}}}} \  A_i^{(c)}(0,t) \ &\leq \ A_e(0,t) \ \stackrel{(a)}{\leq} \ \sum_{\tau=0}^{t}K_{e}(\tau), \label{abc}\\
    &\text{and}\nonumber \\
     \sum_{\mathclap{\substack{(i,c): e\in T_i^{(c)},\\ T_i^{(c)} \in \mathcal{T}^{(c)}}}} \ A_i^{(c)}(0,t) \ &\leq \ A_e(0,t) \ \stackrel{(b)}{\leq} \ \sum_{\tau=0}^{t}\gamma_{e},  \label{abd}
\end{align}
where the left-most sides of the inequalities \eqref{abc} and \eqref{abd} denote the number of delivered packets which has crossed the edge $e$ by the time $t$. This is clearly a lower-bound on $A_e(0,t)$. The inequality (a) in Eqn. \eqref{abc} arises from the fact that the total number of quantum keys generated by the QKD link $e$ up to time $t$ is an upper bound to the number of packets that have crossed the edge till time $t$. Similarly, the inequality (b) in Eqn. \eqref{abd} arises from the fact that the number of packets that have crossed edge $e$ till time $t$ cannot be greater than the cumulative capacity of the link up to time $t$. 

Combining inequalities \eqref{abc} and \eqref{abd}, we have a tighter bound:
\begin{align}
    \sum_{\mathclap{\substack{(i,c): e\in T_i^{(c)},\\ T_i^{(c)} \in \mathcal{T}^{(c)}}}} \  A_i^{(c)}(0,t) \ &\leq \min\Bigg(\sum_{\tau=0}^{t}K_{e}(\tau), \sum_{\tau=0}^{t}\gamma_{e} \Bigg). \label{bbq}
\end{align}

Dividing both sides of the above inequality by $t$ and taking the limit $t \to \infty$ we get
\begin{align} 
    \lim_{t \to \infty} \ \ \sum_{\mathclap{\substack{(i,c): e\in T_i^{(c)},\\ T_i^{(c)} \in \mathcal{T}^{(c)}}}} \ \frac{A_i^{(c)}(0,t)}{t} &\leq \min\Bigg( \lim_{t \to \infty} \frac{\sum_{\tau=0}^{t}K_{e}(\tau)}{t},  \lim_{t \to \infty}\frac{\gamma_{e}t}{t} \Bigg) \label{tgr}
\end{align}
Using Eqn. \eqref{eq:f} on LHS and SLLN on first term in the RHS of Eqn. \eqref{tgr}, we get
\begin{align}
    \ \sum_{\mathclap{\substack{(i,c): e\in T_i^{(c)},\\ T_i^{(c)} \in \mathcal{T}^{(c)}}}} \ \lambda_i^{(c)} \leq \min(\eta_e, \gamma_e) = \omega_e.
\end{align}
From the definition in Eqn. \eqref{rdde}, we see that the condition that no edge is overloaded translates to
\begin{align}
    \lambda_e \leq \omega_e.
\end{align}
This establishes the converse part of Theorem \ref{cap_ch}. $\blacksquare$

\subsection{Throughput-Optimality of TQD} \label{tput}
For any class $c \in \mathcal{C}$, let $A^{(c)}(0,t)$ be the total number of incoming packets belonging to class $c$ up to time $t$. The total number of packets $R^{(c)}(t)$ commonly received by all destination nodes $\mathcal{D}^{(c)}$ of class $c$ can be bounded as follows:
\begin{align}\label{eqn}
  A^{(c)}(0,t) -\sum_{e \in E}X_e(t) -\sum_{e \in E}Y_e(t) \stackrel{(a)}{\leq} R^{(c)}(t) \stackrel{(b)}{\leq}  A^{(c)}(0,t).
\end{align}
Here the first inequality $(a)$ arises from the observation that if a packet $p$ of class $c$ has not reached all destination nodes $\mathcal{D}^{(c)}$, then at least one copy of it must be present in some of the physical queues. Inequality $(b)$ states the obvious fact that the number of packets received till time $t$ is less than the number of packets that have arrived at the source till time $t$.
Since the TQD policy is proven to be rate stable, we know that  
\begin{align*}
  \lim_{t \to \infty}\frac{\sum_{e \in E}X_e(t)}{t} = 0 \quad \text{and} \quad  \lim_{t \to \infty}\frac{\sum_{e \in E}Y_e(t)}{t} = 0.
\end{align*}
Thus, dividing both sides of the inequality \eqref{eqn} by $t$ and taking the limit $t \to \infty,$ we get
\begin{align*}
   \lim_{t \to\infty}\frac{A^{(c)}(0,t)}{t} \leq \lim_{t \to \infty}\frac{R^{(c)}(t)}{t} \leq  \lim_{t \to \infty}\frac{A^{(c)}(0,t)}{t}.
\end{align*}
Thus:
\begin{align*}
  \lim_{t \to \infty}\frac{R^{(c)}(t)}{t} =  \lim_{t \to \infty}\frac{A^{(c)}(0,t)}{t} = \lambda^{(c)}, \forall c \in \mathcal{C}.
\end{align*}
This shows that the TQD policy is secure and throughput optimal. $\blacksquare$

\subsection{Pseudocode for Extended-TQD} \label{e-TQD-section}

\begin{algorithm}
\begin{algorithmic}[1]
\REQUIRE Set of edges $E_S$ with an overlay QKD module, Graph $\mathcal{G}(V,E_S)$, Virtual Queue lengths $\{\tilde{X}_e(t), e \in E \}$ and $\{\tilde{Y}_e(t), e \in E \}$ 
  \item[1:] \textbf{(Edge-Weight Assignment)} Assign each edge  $e \in E_S$ a weight $W_e(t)$ equal to $\tilde{X}_e(t)+\tilde{Y}_e(t)$, i.e.,
$$\boldsymbol{W}(t) \leftarrow\boldsymbol{ \tilde{X}}(t)+\boldsymbol{ \tilde{Y}}(t).$$
  \item[2:] \textbf{(Route Assignment)} Compute a Minimum-Weight Route $T^{(c)}(t) \in \mathcal{T}^{(c)}(t)$ for a class $c$ incoming packet in the \emph{weighted induced graph} $\mathcal{G}(V,E_S)$.
  \item[3:] \textbf{(Key Generation)} Generate symmetric private keys for every edge $e$ via QKD and store them in the key banks.
  \item[4:] \textbf{(Encryption)} Encrypt the data packets waiting in physical queue $X_e$ with the available keys in the key bank and move the encrypted packets to the downstream queue $Y_e$ for every edge $e$.
  \item[5:] \textbf{(Packet Forwarding)} Forward the encrypted physical packets from the queue $Y_e$ to the queue $X_{e'}$ for every edge $e$ according to some packet scheduling policy (ENTO, FIFO etc). Here $e'$ is the next edge in the assigned route of a packet.
  \item[6:] \textbf{(Decryption)} Decrypt the data packets received at physical queue $X_e$ for every edge $e$ using the symmetric key generated earlier via the QKD process. 
  \item[7:] \textbf{(Queue Counter Updation)} Update the virtual key queues and virtual data queues assuming a precedence-relaxed system, i.e.,
  $$ \tilde{X}_e(t+1) \leftarrow \big(\tilde{X}_e(t) + A^{\pi}_e(t) - \kappa_e(t)\big)^{+},\quad \forall e \in E$$
  $$ \tilde{Y}_e(t+1) \leftarrow \big(\tilde{Y}_e(t) + A^{\pi}_e(t) - \gamma_e \big)^{+},\quad \forall e \in E.$$
\end{algorithmic}
\caption{Extended TQD (\emph{e}-TQD) algorithm for packets from the secured sources $\mathcal{S}^*$ requiring quantum encryption}
\label{tqd-algo2}
\end{algorithm}

\newpage

\begin{algorithm}
\begin{algorithmic}[1]
\REQUIRE Graph $\mathcal{G}(V,E)$, Virtual Queue lengths $\{\tilde{X}_e(t), e \in E \}$ and $\{\tilde{Y}_e(t), e \in E \}$ 
  \item[1:] \textbf{(Edge-Weight Assignment)} Assign each edge  $e \in E$ a weight $W_e(t)$ equal to $\tilde{Y}_e(t)$, i.e.,
$$\boldsymbol{W}(t) \leftarrow\boldsymbol{ \tilde{Y}}(t).$$
  \item[2:] \textbf{(Route Assignment)} Compute a Minimum-Weight Route $T^{(c)}(t) \in \mathcal{T}^{(c)}(t)$ for a class $c$ incoming packet in the weighted graph $\mathcal{G}(V,E)$.
%  \item[3:] \textbf{[Key Generation]} Generate symmetric private keys for every edge $e$ via QKD and store them in the key banks.
 % \item[4:] \textbf{[Encryption]} Encrypt the data packets waiting in physical queue $X_e$ with the available keys in the key bank and move the encrypted packets to the downstream queue $Y_e$ for every edge $e$.
  \item[5:] \textbf{(Packet Forwarding)} Forward the physical packets from the queue $Y_e$ to the queue $Y_{e'}$ for every edge $e$ according to some packet scheduling policy (ENTO, FIFO etc). Here $e'$ is the next link in the assigned route of a packet.
 % \item[6:] \textbf{[Decryption]} Decrypt the data packets received at physical queue $X_e$ for every edge $e$ using the symmetric key generated earlier via the QKD process. 
  \item[7:] \textbf{(Queue Counter Updation)} Update the virtual data queues assuming a precedence-relaxed system, i.e.,
  %$$ \tilde{X}_e(t+1) \leftarrow \big(\tilde{X}_e(t) + A^{\pi}_e(t) - \kappa_e(t)\big)^{+},\quad \forall e \in E$$
  $$ \tilde{Y}_e(t+1) \leftarrow \big(\tilde{Y}_e(t) + A^{\pi}_e(t) - \gamma_e \big)^{+},\quad \forall e \in E.$$
\end{algorithmic}
\caption{Extended TQD (\emph{e}-TQD) algorithm for packets from sources ${\mathcal{S}^*}^c$ without needing quantum encryption}
\label{tqd-algo3}
\end{algorithm}

\end{document}